\documentclass{acta}

\usepackage{lscape,epsfig}
\usepackage{amssymb}
\usepackage{amsmath}
\usepackage{url}
\usepackage{color}
\usepackage{xspace}

\newcommand{\mnras}{MNRAS\xspace}
\newcommand{\aap}{A\&A\xspace}

\newcommand{\aj}{AJ\xspace}
\newcommand{\apj}{ApJ\xspace}
\newcommand{\ac}{Astron. \& Comput.\xspace}
\newcommand{\apss}{Ap\&SS\xspace}
\newcommand{\pasp}{PASP\xspace}
\usepackage{natbib}

\SetPages{0}{0}

\SetVol{71}{2021}

\begin{document}

\begin{Titlepage}
\Title{Massive search of spot- and facula-crossing events in 1598 exoplanetary transit
lightcurves}

\Author{R.V.~Baluev$^1$,
E.N.~Sokov$^{2,1}$, I.A.~Sokova$^{2,1}$,
V.Sh.~Shaidulin$^1$, A.V.~Veselova$^1$,
V.N.~Aitov$^3$, G.Sh.~Mitiani$^3$, A.F.~Valeev$^{3,4}$, D.R.~Gadelshin$^3$,
A.G.~Gutaev$^{3,5}$, G.M.~Beskin$^{3,5}$, G.G.~Valyavin$^{3,4,1}$, K.~Antonyuk$^4$,
K.~Barkaoui$^{6,7}$, M.~Gillon$^6$, E.~Jehin$^8$, L.~Delrez$^{6,8}$,
S.~Gu{\dh}mundsson$^9$, H.A.~Dale$^{10}$,
E.~Fern\'andez-Laj\'us$^{11,12}$, R.P.~Di~Sisto$^{11,12}$,
M.~Bretton$^{13}$, A.~Wunsche$^{13}$,
V.-P.~Hentunen$^{14}$, S.~Shadick$^{15}$, Y.~Jongen$^{16}$,
W.~Kang$^{17}$, T.~Kim$^{17,18}$,
E.~Pak\v{s}tien\.e$^{19}$, J.K.T.~Qvam$^{20}$, C.R.~Knight$^{21}$, P.~Guerra$^{22}$,
A.~Marchini$^{23}$, F.~Salvaggio$^{23}$, R.~Papini$^{23}$,
P.~Evans$^{24}$, M.~Salisbury$^{25}$, J.~Garlitz$^{26}$, N.~Esseiva$^{27}$, Y.~Ogmen$^{28}$,
P.~Bosch-Cabot$^{29}$, A.~Selezneva$^{29}$, and T.C.~Hinse$^{30,31}$}
{$^1$Saint Petersburg State University, 7--9 Universitetskaya Emb., St Petersburg 199034, Russia\\
e-mail: r.baluev@spbu.ru\\
$^2$Central Astronomical Observatory at Pulkovo of Russian Academy of Sciences, Pulkovskoje sh. 65/1, St Petersburg 196140, Russia\\
$^3$Special Astrophysical Observatory, Russian Academy of Sciences, Nizhnii Arkhyz, 369167 Russia\\
$^4$Crimean Astrophysical Observatory, Russian Academy of Sciences, Nauchny, 298409 Russia\\
$^5$Kazan (Volga Region) Federal University, Kazan, 420008 Russia\\
$^6$Astrobiology Research Unit, Universit\'e de Li\`ege, Alle\'e du 6 Ao\^ut 19C, B-4000 Li\`ege, Belgium\\
$^7$Oukaimeden Observatory, High Energy Physics and Astrophysics Laboratory, Cadi Ayyad University, Marrakech, Morocco\\
$^8$Space Sciences, Technologies and Astrophysics Research (STAR) Institute, Universit\'e de Li\`ege, All\'ee du 6 Ao\^ut 19C, B-4000 Li\`ege, Belgium\\
$^9$Nes Observatory, Hraunholl 5, 781 Hornafj{\"o}r{\dh}ur, Southeast Iceland\\
$^{10}$Emory University Department of Physics, 400 Dowman Drive, Suite N201, Atlanta, GA 30322\\
$^{11}$Facultad de Ciencias Astron\'omicas y Geof\'isicas - Universidad Nacional de La Plata, Paseo del Bosque S/N - 1900 La Plata,\\ Argentina\\
$^{12}$Instituto de Astrof\'isica de La Plata (CCT La Plata - CONICET/UNLP), Argentina\\
$^{13}$Baronnies Proven\c{c}ales Observatory, Hautes Alpes - Parc Naturel R\'egional des Baronnies Proven\c{c}ales, F-05150 Moydans, France\\
$^{14}$Taurus Hill Observatory, Warkauden Kassiopeia ry., H\"ark\"am\"aentie 88, 79480 Kangaslampi, Finland\\
$^{15}$Physics and Engineering Physics Department, University of Saskatchewan, 116 Science Place, Saskatoon, Saskatchewan,\\ S7N 5E2, Canada\\
$^{16}$Observatoire de Vaison la Romaine, 1075 RD 51, Le Palis, 84110 Vaison-la-Romaine, France\\
$^{17}$National Youth Space Center, Goheung, Jeollanam-do, 59567, S.Korea\\
$^{18}$Department of Astronomy and Space Science, Chungbuk National University, Cheongju-City, 28644, S.Korea\\
$^{19}$Institute of Theoretical Physics and Astronomy, Vilnius University, Sauletekio al. 3, Vilnius 10257, Lithuania\\
$^{20}$Horten Videreg\r{a}ende Skole, Strandpromenaden 33, 3183 Horten, Norway\\
$^{21}$Ngileah Observatory, 144 Kilkern Road, RD 1. Bulls 4894, New Zealand\\
$^{22}$Observatori Astron\`omic Albany\`a, Cam\'i de Bassegoda s/n, 17733 Albany\`a, Spain\\
$^{23}$Astronomical Observatory, DSFTA - University of Siena, Via Roma 56, 53100 - Siena, Italy\\
$^{24}$El Sauce Observatory, Coquimbo Province, Chile\\
$^{25}$School of Physical Sciences, The Open University, Milton Keynes, MK7 6AA, UK\\
$^{26}$AAVSO, Private Observatory, Elgin, OR 97827, USA\\
$^{27}$Observatory Saint Martin, code k27, Amathay Vesigneux, France\\
$^{28}$Green Island Observatory, Code B34, Gecitkale, Famagusta, North Cyprus\\
$^{29}$Observatori Astron\`omic Albany\`a, 17733 Albany\`a, Girona, Spain\\
$^{30}$Institute of Astronomy, Faculty of Physics, Astronomy and Informatics, Nicolaus Copernicus University, Grudziadzka 5, 87-100 Torun, Poland\\
$^{31}$Chungnam National University, Department of Astronomy, Space Science and Geology, Daejeon, South Korea}

\Received{March 15, 2021}
\end{Titlepage}

\Abstract{We developed a dedicated statistical test for a massive detection of spot- and
facula-crossing anomalies in multiple exoplanetary transit lightcurves, based on the
frequentist $p$-value thresholding. This test was used to augment our algorithmic pipeline
for transit lightcurves analysis. It was applied to $1598$ amateur and professional transit
observations of $26$ targets being monitored in the EXPANSION project. We detected $109$
statistically significant candidate events revealing a roughly $2:1$ asymmetry in favor of
spots-crossings over faculae-crossings. Although some candidate anomalies likely appear
non-physical and originate from systematic errors, such asymmetry between negative and
positive events should indicate a physical difference between the frequency of star spots
and faculae. Detected spot-crossing events also reveal positive correlation between their
amplitude and width, possibly owed to spot size correlation. However, the frequency of all
detectable crossing events appears just about a few per cent, so they cannot explain
excessive transit timing noise observed for several targets.}{transiting exoplanets,
transit photometry, starspots, EXPANSION project, TTV}

\section{Introduction}
There is already a long record of starspots studies, including the detection of
spot-crossing events during an exoplanetary transit. Silva (2003) tested this method on
HD~209458, based on its lightcurves analysis, and obtained parameters of its spots (or
groups of spots). It was supposed that such an approach can be used to evaluate some
characteristics of spots, such as size and position, and consecutive transits may provide
information about spots evolution. This method was later applied to several stars with
known exoplanets. For example, Silva-Valio et~al. (2010) studied transit lightcurves of
CoRoT-2, the observed data were fitted using starspots models with different parameters,
such as spot radius, intensity and longitude.

Later on, Tregloan-Reed et~al. (2013) developed a method for modeling the transit and spots
simultaneously and introduced an IDL computer code PRISM and the optimization algorithm
GEMC. Their method was applied to transit light curves of the WASP-19 system and allowed to
calculate the stellar rotation period and the sky-projected obliquity of the system. The
model was later updated in (Tregloan-Reed et~al., 2015) and used for modeling transits in
WASP-6 system. A plenty of other transit modeling routines is available, such as KSint
(Montalto et~al., 2014) or StarSim (Herrero et~al., 2016). From the other side, Southworth
et~al. (2019) applied simply a visual detection of starspots anomalies and noticed that it
was efficient enough for their goals. Mo\v{c}nik et~al. (2017) revealed recurring sequences
of spots in {\it Kepler} data of Qatar-2. This allowed to accurately measure star rotation
period as well as planet-star spin-orbit alignment angle.

Bradshaw and Hartigan (2014) studied the lifetimes of spots on the Sun and other stars,
taking into account their magnetic stellar activity. In particular, for three main-sequence
stars with planets (Kepler-17, CoRoT-2, CoRoT-6), the sizes and lifetimes of spots
resembled scaled values for the Sun. The authors emphasized the importance of combined
usage of the photometric data, Doppler imaging, and analysis of exoplanet transits.

Namekata et~al. (2019) studied the evolution of starspot regions based on the analysis of
local minima of light curves. The lifetimes and emergence and decay rates of the spots were
estimated for more than $50$ star spots on solar-type active stars in {\it Kepler}
database.

Zaleski et~al. (2019) studied differential rotation of the young solar-type star Kepler-71.
Spots and faculae were characterized using transit lightcurves, and these results were
translated into the maps of magnetic activity. The characteristics of lightcurve variations
were determined based on the lightcurve model of Silva (2003), and the authors also
described (very detailedly) the construction of model light curves taking into account
manifestations of star magnetic activity. They applied a pioneer method of using faculae to
estimate the rotation period of a star, and the estimate was consistent with the value
obtained from starspots.

Aronson and Piskunov (2019) presented a model-free method (the transit imaging technique)
for obtaining a map of brightness variations across the disk of a star based on information
from several transit lightcurves. They aimed to produce a large database of stellar spot
coverage. A map of the star brightness distribution without taking into account spots is
obtained by analyzing the median lightcurve for several transits, then synthetic
lightcurves are constructed and compared with observations, on the basis of which the map
is updated.

In (Netto and Valio, 2020) spots on the young solar-type star Kepler-63 are studied. They
applied Silva (2003) method and fitted transit lightcurves, taking into account possible
spot-crossing anomalies. Almost three hundred starspots were characterized, and it was
found that some spots could have existed for at least 75 days. Yet another attempt to study
the starspots evolution was made for Kepler-17 by Namekata et~al. (2020). The authors
claimed that the evolution and location of spots derived from rotational modulations are
significantly different from those derived from in-transit spots. However, with an accuracy
of up to an order of magnitude, their estimates for the rate of emergence and decay of
spots are consistent with similar values for sunspots. The authors therefore suggested the
similarity of the processes of spot formation for solar-type star.

The issue of starspots can also be viewed from another point, namely how they may affect
the best fitting exoplanetary parameters. Czesla et~al. (2009) considered the effect of
starspots and faculae on transit lightcurves and on the normalization of transit profiles.
They redetermined the inclination of the orbit and the radius of the planet in the CoRoT-2
system, taking into account data on the spot activity. This asserts the need to take into
account the effects of stellar activity when obtaining the parameters of exoplanets with an
accuracy of better than a percent level.

There are multiple ways how spots can affect estimations of exoplanetary parameters. Spots
behind a transiting planet lead to an underestimation of its radius, and if these spots are
located near the limb they may cause inaccuracies in transit duration, hence, in orbital
semi-major axis. Near-limb spots can also trigger a spurious transit timing variation
(TTV). Silva-Valio et~al. (2010) considered CoRoT-2 system and showed that spot-crossing
events disturb planet parameters estimates by several percent.

Sanchis-Ojeda et~al. (2011) considered this issue in the context of verifying the
spin--orbit alignment. They used several transits of WASP-4b and analyzed them taking the
effect of starspot occultations. It was claimed that such an approach gives more
constraining result for the sky-projected stellar obliquity than the Rossiter--McLaughlin
method.

Kipping (2012) presented a very detailed description of the model, which takes into account
the differential rotation, non-linear limb darkening, the evolution of spots, and so on.
Their \texttt{macula} code allows to reduce errors in the analysis of photometric data, as
well as to speed-up calculations. Among other effects, the model can take into account the
so-called T$\delta$V, or the gain of the apparent transit depth.

Juvan et~al. (2018) developed PyTranSpot routine that allows to model transit lightcurves
taking into account effects of stellar activity. The technique was merged with the MCMC
method. The authors tested the method on the synthetic lightcurves, and performed the
analysis of WASP-41 system.

As we can see, different researchers agree that spots or faculae appearing along the
transit chord may significantly disturb its parameters and lead to inaccurate conclusions.
Therefore, such anomalies must be detected in each transit lightcurve and fitted. However,
numerous models and codes are available that allows to approximate such spot-crossing
anomalies. Different methods vary from visual perception to quite complicated codes that
take into account multiple effects. However, the detection of spot-crossing anomalies is a
signal detection task, after all. We find that statistical issues related to spots
detection and relevant significance thresholds have not been studied well enough yet.
Without that, it appears difficult to estimate the reliability of numerous individual
results obtained in this domain, and in particular to resolve practical contradictions
about whether a given transit lightcurve demonstrates statistically significant spot
anomalies or not (Baluev et~al., 2020). This becomes increasingly important when we conduct
massive analysis of large number of transits like in (Baluev et~al., 2019). In this work we
present some mathematical results of how to perform a statistically rigorous testing of
spot anomalies. We also construct the corresponding computing pipeline and apply it to our
sample of $\sim 1600$ transit lightcurves.

The paper obeys the following scheme. We discuss transit data that we used, together with a
general overview of their analysis algorithm, in Section~\ref{sec_pipe} We present a
solution to several mathematical and algorithmic issues of spots detection in
Section~\ref{sec_search} Finally, we discuss the results of our spot-search analysis in
Section \ref{sec_res}

\section{Transit data and overview of the full analysis pipeline}
\label{sec_pipe}
We used a moderately expanded update of the data used by Baluev et~al. (2019). Presently we
have $1598$ transit lightcurves for $26$ targets, with $>4\cdot 10^5$ photometric
measurements in total. As before, we use transit photometry from the EXPANSION
(EXoPlanetary trANsit Search with an International Observational Network) project (Sokov
et~al., 2018), which involves a network of amateur and professional observatories. We also
use transit photometry available in published literature, the sources are listed in
Table~\ref{tab:refs}. We did not aim here to construct a comprehensive transit database, so
some objects may possibly miss some known data, especially because not all of them were
updated in 2020-2021.
\begin{table}
\tiny
\caption{Sources of the photometric data (not including the EXPANSION project).}
\label{tab:refs}
\begin{tabular}{llp{80mm}}
\hline
Target    & References              & Note \\
\hline
CoRoT-2   & Gillon et~al. (2010)\\
          & TRAPPIST                & from (Baluev et~al., 2019) \\
\hline
GJ~436    & Gillon et~al. (2007)\\
          & Bean et~al. (2008)      & HST Fine Guidance Sensor\\
          & Shporer et~al. (2009)\\
          & C\'{a}ceres et~al. (2009)       & Very high cadence; we binned these data to $10$~sec chunks\\
          & Christiansen et~al. (2010)  & NASA EPOXI mission \\
\hline
HAT-P-3   & Torres et~al. (2007)\\
          & Chan et~al. (2011)\\
          & Nascimbeni et~al. (2011a)\\
          & Mancini et~al. (2018)\\
\hline
HAT-P-4   & Christiansen et~al. (2010)  & NASA EPOXI mission \\
\hline
HAT-P-12  & Hartman et~al. (2009)\\
          & Lee et~al. (2012)\\
          & Hinse et~al. (2015)\\
          & Sada and Ram{\'o}n-Fox (2016) & These data were kindly provided by the authors\\
          & Mancini et~al. (2018)\\
          & Alexoudi~et~al. (2018)\\
\hline
HAT-P-13  & Bakos et~al. (2009)\\
          & Szab\'{o} et~al. (2010)\\
          & Nascimbeni et~al. (2011b)\\
          & Fulton et~al. (2011)\\
          & Southworth et~al. (2012)\\
          & Sada and Ram{\'o}n-Fox (2016) & These data were kindly provided by the authors\\
\hline
HAT-P-38  & Sato et~al. (2012)\\
\hline
HD~189733 & Bakos et~al. (2006)\\
          & Winn et~al. (2007b)     & T10APT data involve erratic HJD correction (priv.~comm.), we used data kindly provided by the authors\\
          & Pont et~al. (2007)      & HST Advanced Camera for Surveys\\
          & McCullough et~al. (2014) & HST Wide Field Camera 3\\
          & Kasper et~al. (2019)    & Multi-band transmission spectroscopy; very high accuracy data\\
\hline
Kelt-1    & Siverd et~al. (2012)\\
          & Maciejewski et~al. (2018)\\
\hline
Qatar-2   & Bryan et~al. (2012)         & We assumed BJD TDB for the ``BJD'' times.\\
          & Mancini et~al. (2014)\\
\hline
Qatar-4   & Mallonn et~al. (2019) \\
\hline
TrES-1    & Winn et~al. (2007a)\\
\hline
WASP-2    & Southworth et~al. (2010) & Danish telescope clock might have a shift (J.~Southworth, priv. comm.)\\
\hline
WASP-3    & Tripathi et~al. (2010)\\
          & Nascimbeni et~al. (2013)\\
          & Christiansen et~al. (2010)  & NASA EPOXI mission \\
\hline
WASP-4    & Wilson et~al. (2008)\\
          & Gillon et~al. (2009b)\\
          & Winn et~al. (2009)      & Superseded by Sanchis-Ojeda et~al. (2011)\\
          & Southworth et~al. (2009a) & Superseded by Southworth et~al. (2019)\\
          & Sanchis-Ojeda et~al. (2011)\\
          & Nikolov et~al. (2012)\\
          & Petrucci et~al. (2013)  & These data were kindly provided by the authors\\
          & Hoyer et~al. (2013)     & from (Baluev et~al., 2020)\\
          & Huitson et~al. (2017)   & from (Baluev et~al., 2020)\\
          & Southworth et~al. (2019)&\\
          & TRAPPIST                & from (Baluev et~al., 2020)\\
          & TESS                    & from (Baluev et~al., 2020)\\
\hline
WASP-5    & Southworth et~al. (2009b) &\\
          & TRAPPIST                & from (Baluev et~al., 2019) \\
\hline
WASP-6    & Gillon et~al. (2009a) \\
          & Tregloan-Reed et~al. (2015) \\
          & TRAPPIST                & from (Baluev et~al., 2019) \\
\hline
WASP-12   & Hebb et~al. (2009)      & These data were kindly provided by the authors\\
          & Chan et~al. (2011)\\
          & Maciejewski et~al. (2013)& Partly superseded by Maciejewski et~al. (2016)\\
          & Stevenson et~al. (2014)  & Multi-band transmission spectroscopy; very high accuracy data\\
          & Maciejewski et~al. (2016)\\
          & Maciejewski et~al. (2018)\\
\hline
WASP-35   & TRAPPIST\\
\hline
WASP-50   & Gillon et~al. (2011)    & \\
          & Sada et~al. (2012)      & These data were kindly provided by the authors\\
          & Tregloan-Reed and Southworth (2013) & Published lightcurves had an erratic BJD correction; we used correct ones kindly provided by J.~Southworth\\
          & Sada (2018)             & These data were kindly provided by the authors\\
          & TRAPPIST\\
\hline
WASP-52   & Chen et~al. (2017)      & Multi-band transmission spectroscopy; very high accuracy data\\
          & Mancini et~al. (2017)\\
\hline
WASP-75   & G{\'o}mez Maqueo~Chew et~al. (2013)\\
          & TRAPPIST\\
\hline
WASP-84   & Anderson et~al. (2014)\\
          & TRAPPIST\\
\hline
WASP-122  & Turner et~al. (2016) \\
          & TRAPPIST\\
\hline
XO-2N     & Fernandez et~al. (2009)\\
          & Kundurthy et~al. (2013)\\
          & Damasso et~al. (2015)\\
\hline
XO-5      & Burke et~al. (2008)     & \\
          & P{\'a}l et~al. (2009)   & These data were kindly provided by the authors\\
          & Maciejewski et~al. (2011) & Taken from G.~Maciejewski personal web page \\
          & Sada et~al. (2012)      & These data were kindly provided by the authors\\
          & Hinse et~al. (2015)\\
          & Smith (2015)            & These data were kindly provided by the authors\\
          & Kjurkchieva et~al. (2018) & Not clear whether the data are HJD or JD, we assumed HJD UTC\\
\hline
\end{tabular}
\end{table}

In this work we use a reduced version of the pipeline from (Baluev et~al., 2015, 2019), as
implemented in the opensource {\sc PlanetPack} software (Baluev, 2013c, 2018), though
augmented with our search of spot-crossings anomalies. The latter part is described below
(Sect.~\ref{sec_search}), and now we discuss only the basic fittig pipeline.

We run only two fitting stages from (Baluev et~al., 2019). Stage~1 represents an initial
fit used to detect photometric outliers. Now we filtered outliers a bit more aggressively
than in (Baluev et~al., 2019). The threshold was chosen close to $4$-sigma, removing about
$0.05\%$ of individual photometric measurements. This more strict filtering was chosen
because a single outlier may be misinterpreted as a spot anomaly in some cases. The spot
anomalies are detected on Stage~2, which was applied to data already cleaned from outliers.

Each stage involves a maximum-likelihood fit with a dedicated Gaussian Process (GP) model
that remained basically the same as in (Baluev et~al., 2019). We fitted all lightcurves of
a particular target using the same transit parameters bound between lightcurves (these
parameters are planet radius, impact parameter, transit duration). The limb darkening was
modeled using a quadratic law with fittable coefficients, but those coefficients were bound
for all lightcurves belonging to the same or similar spectral band. For example,
lightcurves obtained for a particular target in $R_J$, $R_C$, $r$, or $r'$ filters all
involved the same limb darkening coefficients.

Aside from WASP-12 and WASP-4, which both reveal a quadratic deviation of transit times
(Maciejewski et~al., 2016; Bouma et~al., 2019; Baluev et~al., 2020), no other target in our
list demonstrated statistically significant TTV. Therefore, in this work we also fix
transit times at the quadratic ephemeris (with fittable coefficients). We expect that such
a restriction would make our search of spot-crossing anomalies more reliable for certain
problematic lightcurves.

Each lightcurve also included a cubic polynomial to take into account possible systematic
drifts. Random photometric noise was fitted using a GP model with mandatory white and
optional red component. The white noise was modeled through a fittable jitter term, using
the model from (Baluev 2015a), which is resistant with respect to numeric peculiarities of
the likelihood function. The red noise was modeled through the exponential correlation
function $\exp(-|\Delta t|/\tau)$ with fittable $\tau$. Red noise was first detected in
individual lightcurves as described in (Baluev et~al., 2019), and only robustly fittable
red noise terms were included in the model. After that, we tried to fit red noise in all
the remaining lightcurves under restriction that their $\tau$ is the same, and again left
only those red noise terms that had a robust fit. Lightcurves where the red noise remained
ill-fitted both in the free-$\tau$ and shared-$\tau$ treatment were left with white-only
noise model (such lightcurves would typically imply a negative red noise, meaning blue
noise that we do not consider).

\section{Search of spot-crossing transit anomalies with strict statistical testing}
\label{sec_search}
\subsection{Spot anomalies detection: the statistical theory}
Each spot- or facula-crossing event triggers a bell-like anomaly in the transit curve that
we model by a Gaussian shape:
\begin{equation}
m_{\rm GA}(t,K,\mu,\sigma) = K \exp\left(-\frac{(t-\mu)^2}{2\sigma^2}\right),
\label{GA}
\end{equation}
where $K$ is the amplitude of the signal, $\mu$ being its central time, and $\sigma$ being
characteristic width. Such Gaussian Anomaly (GA) is added to the transit model $m_{\rm
transit}(t,\mathbf p)$, where $m$ means magnitude and $\mathbf p$ is the vector of fittable
transit parameters. Following this convention, $K>0$ for facula-crossings and $K<0$ for
spot-crossings.

Our first task is to detect all statistically significant GAs in a set of the transit
lightcurves. This can be done e.g. by numeric minimization of the $\chi^2$ function
associated to the model $m_{\rm transit}+m_{\rm GA}$. A bit more general approach taking
into account poorly known noise level is to also use a parameterized noise model and to
maximize the corresponding likelihood function (Baluev, 2009). The latter approach can be
easily extended to treat the correlated photometric noise via the GP model (Baluev, 2011;
Baluev, 2013b; Rajpaul et~al., 2015; Foreman-Mackey et~al., 2017; Angus et~al., 2018).

However, when fitting nonlinear models like~(\ref{GA}) we have to solve a computationally
complicated optimization task. This task is made so heavy because the likelihood function
typically has multiple peaks corresponding to different positions in the plane of nonlinear
parameters $(\mu,\sigma)$. Each such local maximum of the likelihood corresponds to a
single local solution for~(\ref{GA}), and different such solutions appear nearly
uncorrelated in terms of their best fitting parameters. From a mathematical point of view,
the cause of such a behavior comes from the following correlation measure:
\begin{equation}
{\rm corr}(\mu_{\rm GA}^{(1)},\mu_{\rm GA}^{(2)}) = \frac{\left|\int\limits_{-\infty}^{+\infty} \mu_{\rm GA}^{(1)}(t) \mu_{\rm GA}^{(2)}(t) dt\right|}{\sqrt{\int\limits_{-\infty}^{+\infty} \left[\mu_{\rm GA}^{(1)}(t)\right]^2 dt \int\limits_{-\infty}^{+\infty} \left[\mu_{\rm GA}^{(2)}(t)\right]^2 dt}} = \sqrt{\frac{2\frac{\sigma_2}{\sigma_1}}{1+\left(\frac{\sigma_2}{\sigma_1}\right)^2}}\, {\rm e}^{-\frac{(\mu_2-\mu_1)^2}{2(\sigma_1^2+\sigma_2^2)}}.
\label{GAcorr}
\end{equation}
We can see that it decreases for large $|\mu_2-\mu_1|$ or for large
$|\log(\sigma_2/\sigma_1)|$. Therefore, if either $\mu_{1,2}$ or $\sigma_{1,2}$ differ too
much, the two models $\mu_{\rm GA}^{(1)}$ and $\mu_{\rm GA}^{(2)}$ can be treated as
(quasi-)independent ones even though they both are expressed by formally the same
function~(\ref{GA}). Then the entire plane $(\mu,\sigma)$ is split into a set of
``independence cells'' such that correlations between different models $\mu_{\rm GA}$ are
high within a single cell, while distinct cells are only weakly correlated in average. Then
total number of local maxima of the $\chi^2$ (or likelihood) function is roughly equal to
the number of such cells, and each cell would typically contain just a single maximum.
Notice that the amplitude $K$ is a linear parameter, so it cannot generate quasi-independent
models: the correlation~(\ref{GAcorr}) does not depend on $K_{1,2}$. Hence, for each $\mu$
and $\sigma$ there is only a single best fitting value of $K$.

The effect of multiple likelihood peaks owed to nonlinear parameters is explained in more
details in (Baluev, 2013a, 2015b). We cannot know in advance which peaks would appear high
or low, and we do not have restrictive enough prior information about possible parameters
of GAs. So we have to directly scan some reasonable domain in the $(\mu,\sigma)$ plane
seeking the highest peak (the global maximum inside domain). In other words, we should test
multiple candidate solutions~(\ref{GA}), starting each fit from a point inside a separate
independence cell.

A quite similar phenomenon is known for periodograms, which can be viewed in a direct
relationship with the least squares and maximum likelihood fitting (Lomb, 1976; Scargle,
1982; Baluev, 2014). In this case multiple peaks appear because of the nonlinear frequency
parameter $f$. The width of each ``independence cell'' in the frequency axis is about
$\Delta f \sim 1/T$, where $T$ being the time series length. Two sinusoidal variations that
have $|f_2-f_1|\gtrsim 1/T$ appear independent in terms of the correlation measure
analogous to~(\ref{GAcorr}). This effect simply determines the periodogram resolution: we
have to scan the periodogram with the step $\sim 1/T$ at largest, or we may undersample (or
even miss) the global maximum. So we have to perform numerous independent fits to determine
just a single nonlinear parameter, the frequency.

However, the issues coming from nonlinear parameters are more important than just the
increased computing load. An important caveat is that by such wide scanning we implicitly
test a large number of statistically independent solutions, each corresponding to a single
likelihood peak. Since all such solutions are nearly independent statistically, this leads
to an increased false positives rate. This is owed to the statistical effect of multiple
testing: to make a mistake with, say, $1000$ peaks tested at once is roughly $1000$ times
more probable, compared to a single test. This effect significantly increases all the
detection thresholds. In the periodograms theory this is well known as the ``bandwidth
penalty''. In our task of GA detection a similar effect should appear, even if we test just
a single lightcurve.

The general theory of how to treat this effect for an arbitrary nonlinear signal is given
in (Baluev, 2013a). Mathematically, that theory was considered with periodograms and
periodic signals in mind, hence all formulae include a mandatory frequency parameter. But
this assumption was not critical, so all formulae can be easily promoted to non-periodic GA
models like~(\ref{GA}).

In our case the null model is $m_{\rm transit}(t)$, and the signal is expressed by $m_{\rm
GA}(t)$. We should perform two fits: for just $m_{\rm transit}$ and for $m_{\rm
transit}+m_{\rm GA}$, assuming the parameters $\mu$ and $\sigma$ to be constant. Given
these fits, we can construct the logarithm of the likelihood ratio, $\zeta$, which is a
function of our two nonlinear unknowns $\mu$ and $\sigma$. The maximum of
$\zeta(\mu,\sigma)$ shall determine (via its location) the best fitting values for these
arguments. Notice that $\mu$ and $\sigma$ are treated separately because of their
nonlinearity, while the remaining parameters are either strictly linear (like $K$) or can
be linearized approximately about the best fitting point (like $\mathbf p$).

Since input data involve noise, $\zeta(\mu,\sigma)$ is a random field, while its global
maximum (it has to be computed numerically) is random quantity. Large
$\max\zeta(\mu,\sigma)$ indicates that our lightcurve cannot be explained well by just
$m_{\rm transit}(t)$ and likely also involves a GA~(\ref{GA}), while small value means that
GAs are unlikely to exist and $m_{\rm transit}(t)$ has satisfactory accuracy. To derive the
detection threshold separating these two decisions, we should statistically quantify the
levels of $\max\zeta$ under the null hypothesis (no GAs).

For that, we should compute the False Alarm Probability (FAP) function, which is
complementary to the distribution function of $\max\zeta$:
\begin{equation}
{\rm FAP}(z) = {\rm Pr}\{ \max\zeta > z \} = 1 - P_{\max}(z), \quad P_{\max}(z) = {\rm Pr}\{ \forall \zeta < z \}.
\end{equation}
The computation of ${\rm FAP}(z)$ is one of primary results in (Baluev, 2013a). For the
dimension two (two nonlinear parameters $\mu,\sigma$) its approximation looks like
\begin{equation}
{\rm FAP}_2(z) \simeq 2 A_2 e^{-z} \sqrt z, \quad A_2 = \frac{1}{2\pi^{\frac{3}{2}}} \int\limits_{\mathcal D} \sqrt{\det {\rm var}(\eta')}\, d\mu\, d\sigma,
\label{FAP2}
\end{equation}
where $\mathcal D$ is the domain in the $(\mu,\sigma)$ plane that we scan for possible GAs,
and ${\rm var(\eta')}$ is the $2\times 2$ variance-covariance matrix of the gradient of an
auxiliary Gaussian random field $\eta$ defined as $\eta(\mu,\sigma)=\pm
\sqrt{2\zeta(\mu,\sigma)}$, with the sign taken from the best fitting $K$.

In~(\ref{FAP2}) we removed an additional correction term responsible for the boundary
maxima, when the global maxima is attained on the boundary of $\mathcal D$. This is because
in our algorithm such cases are treated unreliable (see below) and are eliminated from the
investigation, and so they cannot generate false alarms.

Now, given a small detection threshold $\rm FAP^*$ we may claim that our lightcurve reveals
a statistically significant GA, if ${\rm FAP}(\max\zeta)<{\rm FAP}^*$ for the particular
$\max\zeta$ computed from the actual data. The best fitting GA parameters are then given by
the position of $\max\zeta$. Otherwise, if ${\rm FAP}(\max\zeta)>{\rm FAP}^*$, the
lightcurve is consistent with a clean transit.

The formula~(\ref{FAP2}) refers to a 2D domain in the $(\mu,\sigma)$ plane. However, 2D
scan may appear computationally hard, and we may replace it by a 1D one, in which we fix
$\sigma$ at a reasonable prior value. Such a simplification is justified below, but here we
can give a 1D version of~(\ref{FAP2}) for this case:
\begin{equation}
{\rm FAP}_1(z) \simeq 2 A_1 e^{-z}, \quad A_1 = \frac{1}{2\pi} \int\limits_{T_1}^{T_2} \sqrt{{\rm var}(\eta'_\mu)}\, d\mu,
\label{FAP1}
\end{equation}
In this formula $\sigma$ is assumed constant, so we integrate only over $\mu$.

The coefficient $A$ in~(\ref{FAP2},\ref{FAP1}) is responsible for the penalty of multiple
peaks testing. It is not obvious yet and still needs to be computed. In Sect.~4 of (Baluev,
2013a), expressions of two types were considered: precise formulae (slow) and analytic
approximations (fast). The latter ones were derived assuming a periodic signal in place of
our GA, so they need to be promoted to conditions of our task. The analytic approach is
based on the so-called approximation of ``uniform phase coverage'', where various
summations of periodic functions over the discrete time series are replaced by analytic
integrals over a single period. This cannot be used in our task directly, because the GA
signal~(\ref{GA}) is non-periodic, but we can apply an equivalent approximation. Namely, we
can replace the necessary summations by integrals over the entire time span, assuming that
observations come with a constant cadence and the time span is large. Then, using
so-modified formulae of Sect.~4.1 of (Baluev, 2013a), we obtained the following:
\begin{equation}
{\rm var}(\eta') \simeq \left( \begin{array}{cc}\frac{1}{2\sigma^2} & 0 \\ 0 & \frac{1}{2\sigma^2} \end{array} \right), \quad
\sqrt{\det {\rm var}(\eta')} = \frac{1}{2\sigma^2}, \quad
\sqrt{{\rm var}(\eta'_{\mu})} = \frac{1}{\sigma\sqrt 2}.
\label{detG}
\end{equation}
Let us define the parametric domain $\mathcal D$ as a rectangle with $\mu \in [T_1, T_2]$
(typically, the transit duration range) and $\sigma \in [\Sigma_1,\Sigma_2]$, and then
\begin{equation}
A_2 = \frac{T_2-T_1}{2\pi^{\frac{3}{2}}} \left( \frac{1}{\Sigma_1} - \frac{1}{\Sigma_2} \right), \quad
A_1 = \frac{T_2-T_1}{2\pi\sigma\sqrt 2}.
\label{Afast}
\end{equation}
Notice that~(\ref{detG}) and~(\ref{Afast}) require $\Sigma_{1,2} \ll T_2-T_1$. This allowed
to make several simplifications, in particular by neglecting the correlation between GA and
the null model, ${\rm corr}(m_{\rm transit},m_{\rm GA})$. These approximations also do not
involve correlated noise models (noise is assumed to be white).

As we can see, there are many assumptions and hence multiple potential vulnerabilities with
the approximation~(\ref{Afast}), but it remains not obvious how accurate it can be until we
compare it with a better assessment.

More accurate formula for $A$ comes from the matrix decompositions of Sect.~4.2 of (Baluev,
2013a), namely we use adapted versions of eqs.~(42-44) from that paper. First, rewrite our
GA as $m_{\rm GA} = K g(t,\mu,\sigma)$ and determine the full Fisher information matrix of
our compound model $m_{\rm transit}(t)+m_{\rm GA}(t)$:
\begin{equation}
Q = \left( \begin{array}{cccc}
\left\langle \frac{\partial m_{\rm transit}}{\partial \mathbf p}\otimes \frac{\partial m_{\rm transit}}{\partial \mathbf p} \right\rangle & \left\langle g \frac{\partial m_{\rm transit}}{\partial \mathbf p} \right\rangle & \left\langle \frac{\partial g}{\partial\mu} \frac{\partial m_{\rm transit}}{\partial \mathbf p} \right\rangle & \left\langle \frac{\partial g}{\partial\sigma} \frac{\partial m_{\rm transit}}{\partial \mathbf p} \right\rangle \\
\left\langle \left(\frac{\partial m_{\rm transit}}{\partial \mathbf p}\right)^{\rm T} g \right\rangle & \left\langle g^2 \right\rangle & \left\langle \frac{\partial g}{\partial\mu} g \right\rangle & \left\langle \frac{\partial g}{\partial\sigma} g \right\rangle \\
\left\langle \left(\frac{\partial m_{\rm transit}}{\partial \mathbf p}\right)^{\rm T} \frac{\partial g}{\partial\mu} \right\rangle & \left\langle g \frac{\partial g}{\partial\mu} \right\rangle & \left\langle \left(\frac{\partial g}{\partial\mu}\right)^2 \right\rangle & \left\langle \frac{\partial g}{\partial\sigma} \frac{\partial g}{\partial\mu} \right\rangle \\
\left\langle \left(\frac{\partial m_{\rm transit}}{\partial \mathbf p}\right)^{\rm T} \frac{\partial g}{\partial\sigma} \right\rangle & \left\langle g \frac{\partial g}{\partial\sigma} \right\rangle & \left\langle \frac{\partial g}{\partial\mu} \frac{\partial g}{\partial\sigma} \right\rangle & \left\langle \left(\frac{\partial g}{\partial\sigma}\right)^2 \right\rangle
\end{array} \right).
\label{mQ}
\end{equation}
Here triangular brackets designate the weighted summation over the time series
(substituting the best fitting $\mathbf p$ from the null model).

After that we should compute the Cholesky decomposition $Q = L L^{\rm T}$, where the
low-triangular matrix $L$ would look like:
\begin{equation}
L = \left( \begin{array}{cccc}
L_{{\rm null}, {\rm null}} & 0 & 0 & 0 \\
L_{K, {\rm null}} & l_{K K} & 0 & 0 \\
L_{\mu, {\rm null}} & l_{\mu K} & l_{\mu \mu} & 0 \\
L_{\sigma, {\rm null}} & l_{\sigma K} & l_{\sigma \mu} & l_{\sigma\sigma}
\end{array} \right)
\label{mL}
\end{equation}
Now we need only the diagonal elements of the bottom-right square block, $l_{KK}$,
$l_{\mu\mu}$, and $l_{\sigma\sigma}$. Using them,
\begin{equation}
\sqrt{\det {\rm var}(\eta')} = \frac{l_{\mu\mu} l_{\sigma\sigma}}{l_{KK}^2}, \quad
\sqrt{{\rm var}(\eta'_{\mu})} = \frac{l_{\mu\mu}}{l_{KK}}.
\label{Aslow}
\end{equation}
These quantities can be further integrated numerically using second formula
of~(\ref{FAP2},\ref{FAP1}), and so we obtain $A$.

Even this way of computing the ${\rm FAP}$ is not entirely precise, because we still used
several hidden simplifying assumptions: (i) the noise is still white, (ii) we use pure
least-squares fitting, i.e. there is no fittable noise (noise is known), (iii) original
models from (Baluev, 2013a) assumed strictly linear $m_{\rm transit}$, so we performed its
hidden linearization with respect to $\mathbf p$ in the vicinity of the best fitting
points. The last two issues were already discussed in (Baluev, 2013a) and they are likely
negligible, if our models are not ill-fitted. For example, the ${\rm FAP}$ formulae simply
become ``more approximate'' but still valid, if in place of pure least squares we apply the
maximum-likelihood method with a fittable noise. This is because the corresponding Fisher
information matrix have zeros in the offdiagonal blocks responsible for correlations
between the noise and curve parameters (Baluev, 2009). The linearization of $m_{\rm
transit}$ about the best fitting null model also should not break resulting approximations,
if the fit is robust. However, the first issue (correlated noise) is important because red
photometric noise is quite typical.

Correlated noise models were not considered in (Baluev, 2013a), but the necessary formulae
are not hard to obtain by a minor modification. We need to recompute the Fisher information
matrix~(\ref{mQ}) for the general likelihood function involving a GP noise model (Baluev,
2013b). It appears that we simply need to replace the time-series summation operation
$\langle * \rangle$ in~(\ref{mQ}) by the following bi-linear form:
\begin{equation}
\langle x y \rangle \longmapsto {\mathbf x}^{\rm T} V^{-1} {\mathbf y},
\end{equation}
where $V$ is the covariance matrix of the noise (at the best fitting null model). Notice
that this bi-linear form can be computed faster, profiting from the Cholesky decomposition
of $V$:
\begin{equation}
\langle x y \rangle \longmapsto {\mathbf x}^{\rm T} V^{-1} {\mathbf y} = (L_V^{-1} {\mathbf x})^{\rm T} (L_V^{-1} {\mathbf y}), \quad V = L_V L_V^{\rm T}
\label{bilin}
\end{equation}
The rest of the computation remains the same.

Summarizing all the above, a GA candidate can be tested for statistical significance
using~(\ref{FAP2}) or~(\ref{FAP1}) and: (i) a fast entirely analytic formula~(\ref{Afast}),
(ii) slow but more accurate formulae~(\ref{mQ},\ref{mL},\ref{Aslow}) that still assume only
white noise, (iii) even more slow version of the last set, augmented by~(\ref{bilin}) to
take the red noise into account.

We find that~(\ref{Afast}) is not very accurate in practice. The approximation~(\ref{detG})
has satisfactory accuracy only in the middle of a transit, where $m_{\rm transit}$ varies
slowly. In the ingress or egress phases $m_{\rm transit}$ varies faster, so that its
correlation with $m_{\rm GA}$ is not negligible. The value of $A_1$ computed
using~(\ref{Aslow}) is typically $30-50\%$ larger than the analytic value~(\ref{Afast}).
The red noise, if present, also triggers an increase of $A_1$, depending on the parameters.
In our practical computations numeric values for $A_1$ appeared mostly in the range from
$2$ to $5$. This means that typically we should have about a few or ten likelihood peaks
per each lightcurve. This penalty is not as large as the periodogram bandwidth penalty, but
still it is a big factor that cannot be neglected.

Larger $A$ means larger detection threshold (less number of GAs pass the test). That is,
using undervalued $A$ would lead us to excessive number of false GA detections, so we did
not use the fast formula~(\ref{Afast}) for actual GA testing. However, (\ref{Afast}) and
associated expressions are useful to understand our task better. For example, since
elements of the matrix~(\ref{detG}) are basically variances of the likelihood function
gradient, their inverse values estimate average width of likelihood peaks. This width
appears $\sigma\sqrt 2$ both in $\mu$ and in $\sigma$ variable. This information can be
used to construct a scan grid with an optimal resolution.

\subsection{Spot anomalies search and verification: practical aspects}
When we started to test the method of GA detection in practical lightcurve data, several
additional issues appeared. We highlight three of them: (i) slow computing speed, (ii)
various subtle model inaccuracies and noisy drifts have a tendency to trigger detection of
highly-correlated GAs, rendering the transit model nearly degenerate, (iii) it appeared
difficult to disentangle red noise from GAs.

Concerning the issue of slow computation, it cannot be avoided completely, because we have
to test all probe GAs located in distinct independence cells of the $(\mu,\sigma)$ plane.
However, the speed can be improved if we could replace the full 2D scan by a 1D one with
fixed $\sigma$. Such a replace appears justified by the following explanation.

Let us first estimate typical practical range for the spot-crossing duration (the $\sigma$
parameter). This range depends, primarily, on the statistical distribution of the spot
impact parameter $s$ defined as the distance between planet trajectory and spot, divided by
the planet radius $r$. The widest GAs would appear when planet crosses a spot by its
equator ($s=0$), while ``grazing'' spot-crossings ($s$ close to $1$) would generate GAs
with small $\sigma$. In theory, GAs may have arbitrarily small width, but too narrow
spot-crossings (i) are statistically rare, (ii) are difficult to detect due to a small
amplitude. The quantity $s$ is distributed uniformly in the $[0,1]$ range, so its median
value is $\frac{1}{2}$, while $90\%$ of events occur for $s<0.9$. From the other side, if
the planet disk is circular (and spot itself is small) then the spot-crossing half-duration
is $\tau_{\rm spot} = \sqrt{1-s^2} \tau_{\rm pl}$, where $\tau_{\rm pl}$ is time that
planet takes to pass its radius. Therefore, the median half-duration of a spot-crossing is
$0.87 \tau_{\rm pl}$, while $90\%$ of cases have half-duration longer than $0.44 \tau_{\rm
pl}$. As we can see, narrow events are statistically rare, with only $\sim 13\%$
occurrences below half of the median duration.

As well, physical spot-crossings should not be very wide. Small spots cannot generate events
lasting longer than $2\tau_{\rm pl}$. If a spot (or spot group) is big, compared to the
planet, then the event may last somewhat longer, but spot-crossings wider than e.g. twice of
the above value are not very likely.

Based on these considerations, we adopted the following reasonable range for a
spot-crossing width: $|\log(\sigma/\sigma_{\rm c})| \leq \rho$, where $\rho= \log 2 \approx
0.7$ and $\sigma_{\rm c}$ is some central value. Basically, this range is from $\sigma_{\rm
c}/2$ to $2\sigma_{\rm c}$. The value $\sigma_{\rm c}$ should correspond to the median
half-duration, $0.87 \tau_{\rm pl}$. Notice, however, that we cannot just equate these two
quantities here, because we approximate spot-crossing by a Gaussian shape~(\ref{GA}), while
the actual anomaly should look more box-like (if the spot is not large). The best fitting
value of $\sigma_{\rm c}$ can be obtained by maximizing the correlation~(\ref{GAcorr})
between the GA and box-shaped anomaly with a given half-duration $\tau_{\rm spot}$. By
performing this maximization numerically, we obtained that $\sigma\approx 0.7\tau_{\rm
spot}$, and hence $\sigma_{\rm c}\approx 0.6 \tau_{\rm pl}$.

Given such $\sigma$-range for spot-crossings, its logscale width appears $2\rho\approx 1.4$.
Simultaneously, we already know that likelihood peaks should have typical width of
$\sigma\sqrt 2$ along the $\sigma$ axis. This corresponds to the width of $\sqrt 2$ in
$\log\sigma$, nearly the same as our $\sigma$-range. Therefore, such $\sigma$-range may
embed only a single likelihood peak, and we do not need to formally scan this range. It is
quite safe to simplify our task by scanning only along the $\mu$ parameter (within the
transit range), fixing $\sigma=\sigma_{\rm c}$. We first perform such initial scan to
detect preliminary GA candidates using our 1D criterion (${\rm FAP}_1(z)$), and after that
all the detected GAs are refitted using free $\sigma$. After this fit, to ensure that all
GAs would be detected in the full 2D scan as well, we re-verify them based on the 2D
criterion (${\rm FAP}_2(z)$). Moreover, the coefficient $A_2$ is also computed using
simplifications, in order to avoid direct 2D integration. We first compute $A_1$ using
accurate formulae~(\ref{mQ},\ref{mL},\ref{Aslow}) and then correct this estimate by the
factor $\frac{A_2}{A_1} = 2\sqrt{\frac{2}{\pi}} \sinh\rho \approx 1.2$ that follows from
solely analytic formulae~(\ref{Afast}).

The second issue appears when our algorithm tries to use a GA model to fit something not
suitable. This includes attempts to fit long-term drifts by a near-degenerate superposition
of Gaussians. Such trends are already modeled by cubic polynomials and using a red noise
GP model. Even if these models appear partly inaccurate, it is inadequate to use GA
shapes~(\ref{GA}) for fitting any residual longer-term variation, as this leads to
degeneracy issues and may lead to an over-fit effect. To reject such cases we verify that
each our GA satisfies reliability criteria: (i) after full 2D fit our GA remains within the
domain $\mathcal D$: inside the transit range in terms of $\mu$ and inside the required
$\sigma$-range, (ii) the value of $\sigma$ is smaller than $0.1T$, with $T$ being the
lightcurve time span, (iii) all GAs detected in the same lightcurve must have small enough
correlations~(\ref{GAcorr}), namely smaller than $\frac{1}{3}$. If some GA failed any of
these reliability criteria, it was removed and the model was refit with remaining GAs which
were re-verified anew. Last detected GAs were removed first. The lightcurve with one or more
unreliable GA was no longer tested for more GAs (in practice all such lightcurves
demonstrated weird noisy variations that were fitted as large-magnitude correlated noise).
Notice that our domain tests are applied taking into account uncertainties in $\mu$ and
$\sigma$, that is we keep GAs which nominal values are formally out of the domain, but the
domain still intersects with the uncertainty ranges.

The third issue is to disentangle GAs and correlated noise. Red or quasiperiodic noise
often demonstrate long-living variations that can be represented through a superposition of
GAs. However, if fitting the red noise through a GP model typically increases uncertainties
in other parameters, fitting multiple GAs often causes an overfit effect with undervalued
uncertainties, owed to a mock reduction of the residuals r.m.s. Therefore, it is important
to avoid erratic interpretation of correlated noise through GAs. But the opposite
misinterpretation is also undesired, since our goal is to detect spot-induced GAs, after
all. Moreover, cases may exist where we cannot statistically distunguish these two
interpretations, ``white noise + GA'' or ``red noise without GA''. This ambiguity is
difficult to resolve in any other way but through a prior prioritization of the models.

We adopt such an algorithm that resolves this ambiguity in favour of the GA model. However,
if at any step the GA term appears suspicious, e.g. statistically insignificant or
non-trusted due to strange values of parameters, we fallback to the red noise model without
this GA. Thanks to such a behavior we do not fit lightcurves using GAs if the GA model
itself does not look well justified. The entire algorithm is as follows.
\begin{enumerate}
\item Perform initial 1D detection of GA candidates, filtering away unreliable ones and
assuming only white noise. This would likely produce a somewhat excessive list of GAs.
\item Run the red noise detection algorithm (Baluev et~al., 2019) as detailed above, but also
taking into account all preliminary detected GAs.
\item Retest the GAs in the 2D framework, also filtering away unreliable ones (GAs
parameters might change so some of them may no longer pass the reliability tests), and
using the red noise GP model. Many of the GA candidates do not survive this stage.
\item Those lightcurves where we removed a GA should be retested for possible red noise
again (it will likely appear fittable if it was not fittable before). This assumes a
return to step~2.
\item Steps 2--4 are iterated in a loop until the solution is stabilized. In the end we
have all red noise terms robustly fittable and all GA candidates statistically significant
and passing the reliability tests.
\end{enumerate}

\section{Results}
\label{sec_res}
\begin{figure}[t]
\includegraphics[width=\textwidth]{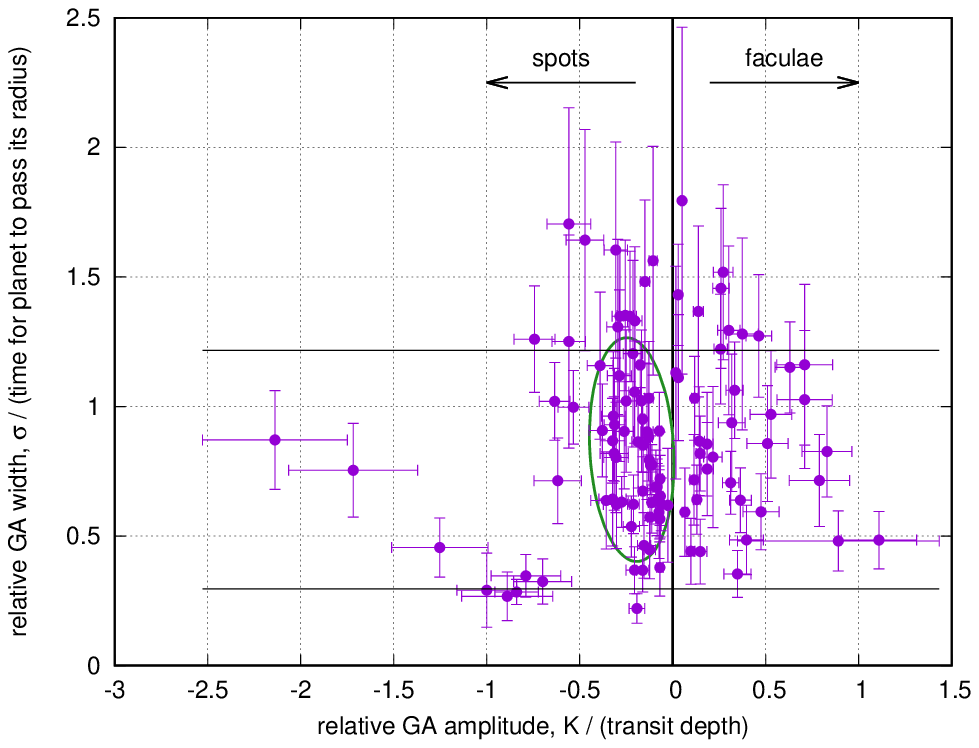}\\
\FigCap{\label{fig:allspots} All detected GAs in the amplitude--width diagram. We outline a
slightly inclined concentration of negative GAs potentially reflecting physical
spot-crossing events. Two horizontal lines label the range where a GA should reside (taking
into account its uncertainties) to pass the reliability test.}
\end{figure}

In $1598$ transit lightcurves our analysis pipeline detected $109$ potential GAs. All these
GA candidates are shown in Fig.~\ref{fig:allspots}, in the form of a 2D diagram ``amplitude
-- width''. We assumed the ${\rm FAP}$ threshold of $0.0027$, which means, formally, that
we should have about $4$ statistical false positives in total. However, one should bear in
mind that this estimate refers to particular adopted models and involves various hidden
assumptions about photometric noise.

\begin{figure}[t]
\includegraphics[width=\textwidth]{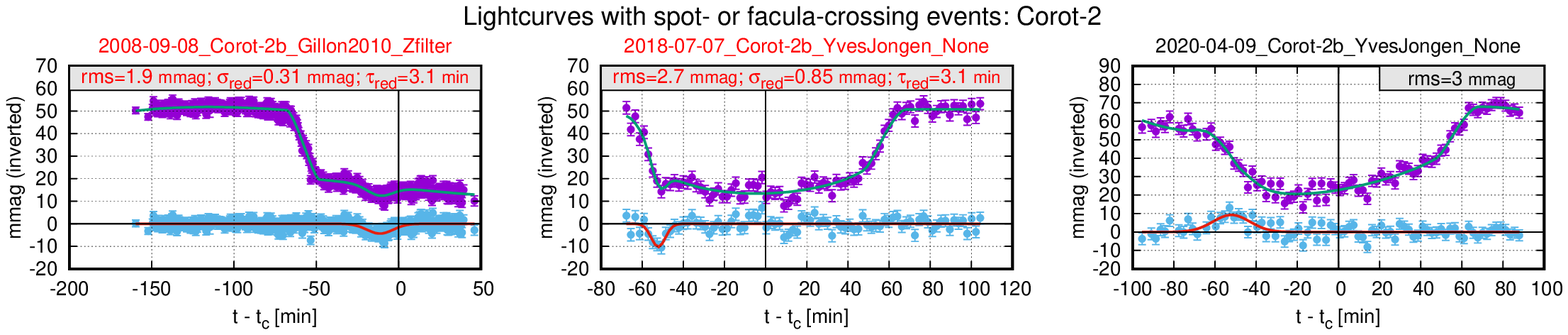}\\
\includegraphics[width=\textwidth]{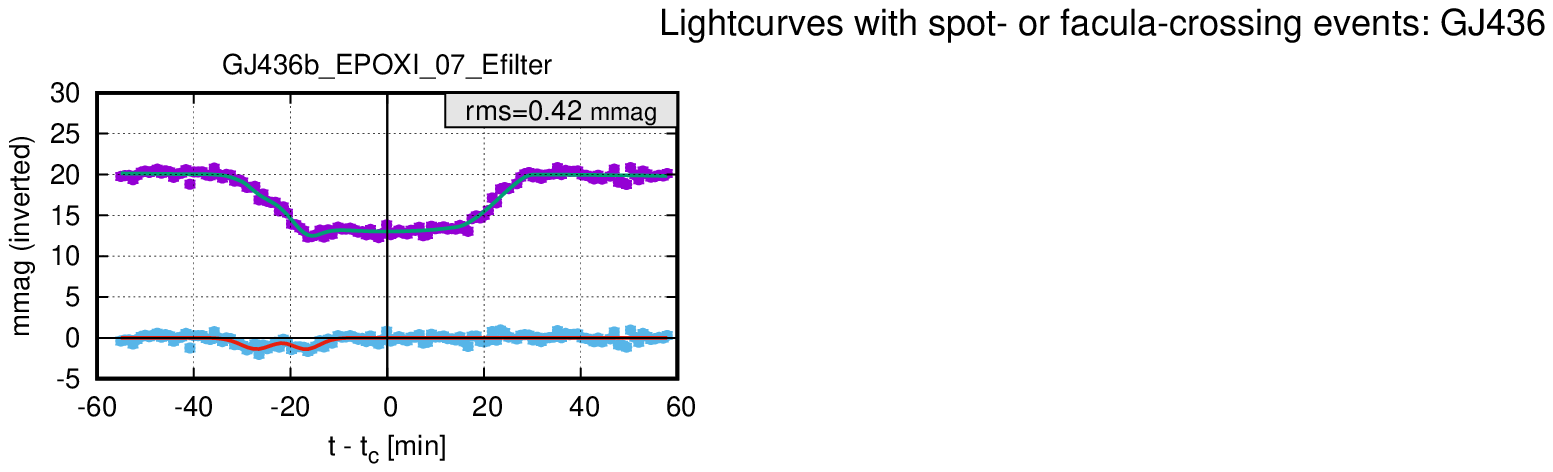}\\
\includegraphics[width=\textwidth]{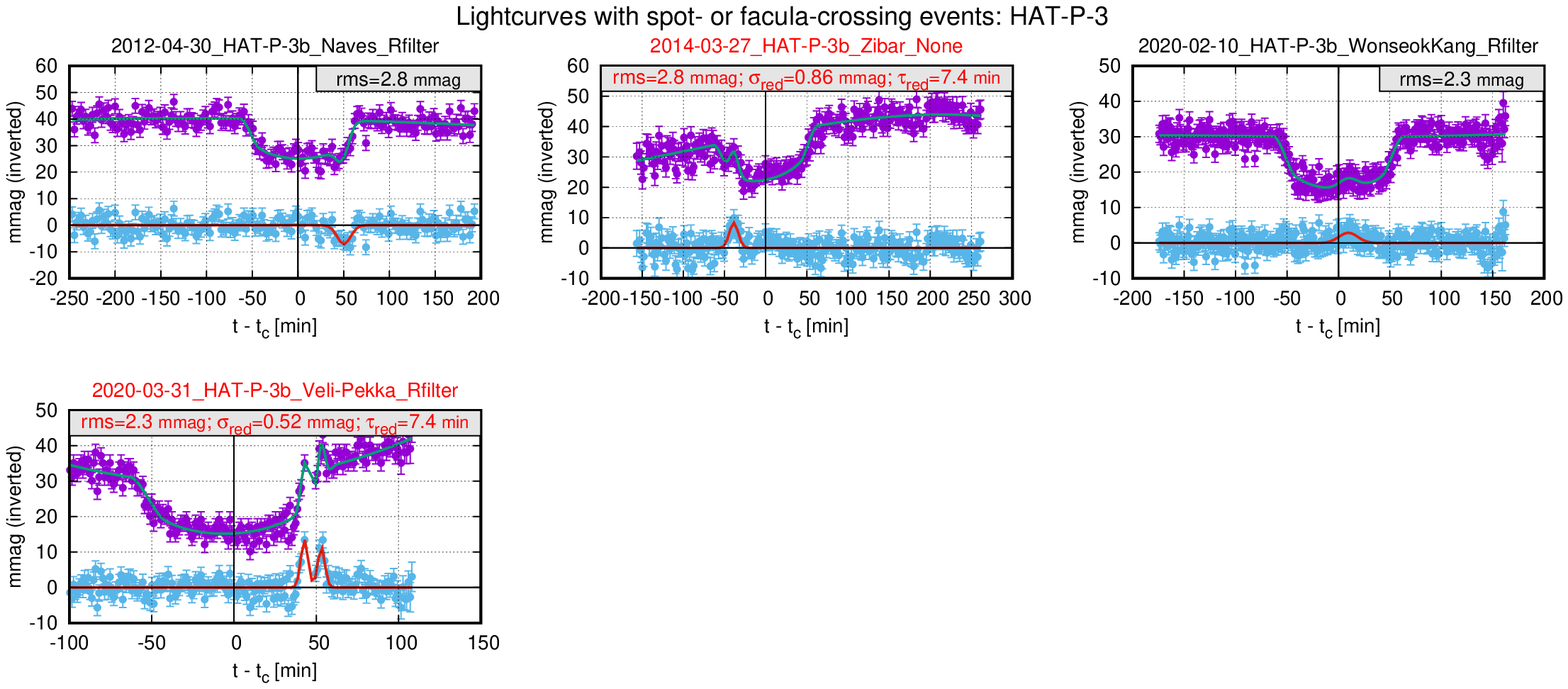}\\
\includegraphics[width=\textwidth]{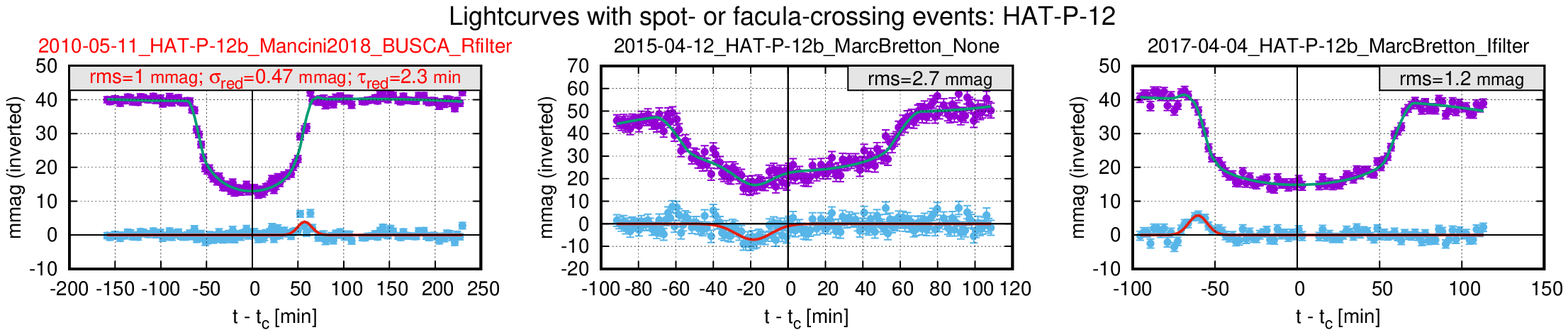}\\
\FigCap{\label{fig:spots1} GAs detected by our pipeline (Part~1: CoRoT-2, GJ~436, HAT-P-3, and HAT-P-12).}
\end{figure}

\begin{figure}[t]
\includegraphics[width=\textwidth]{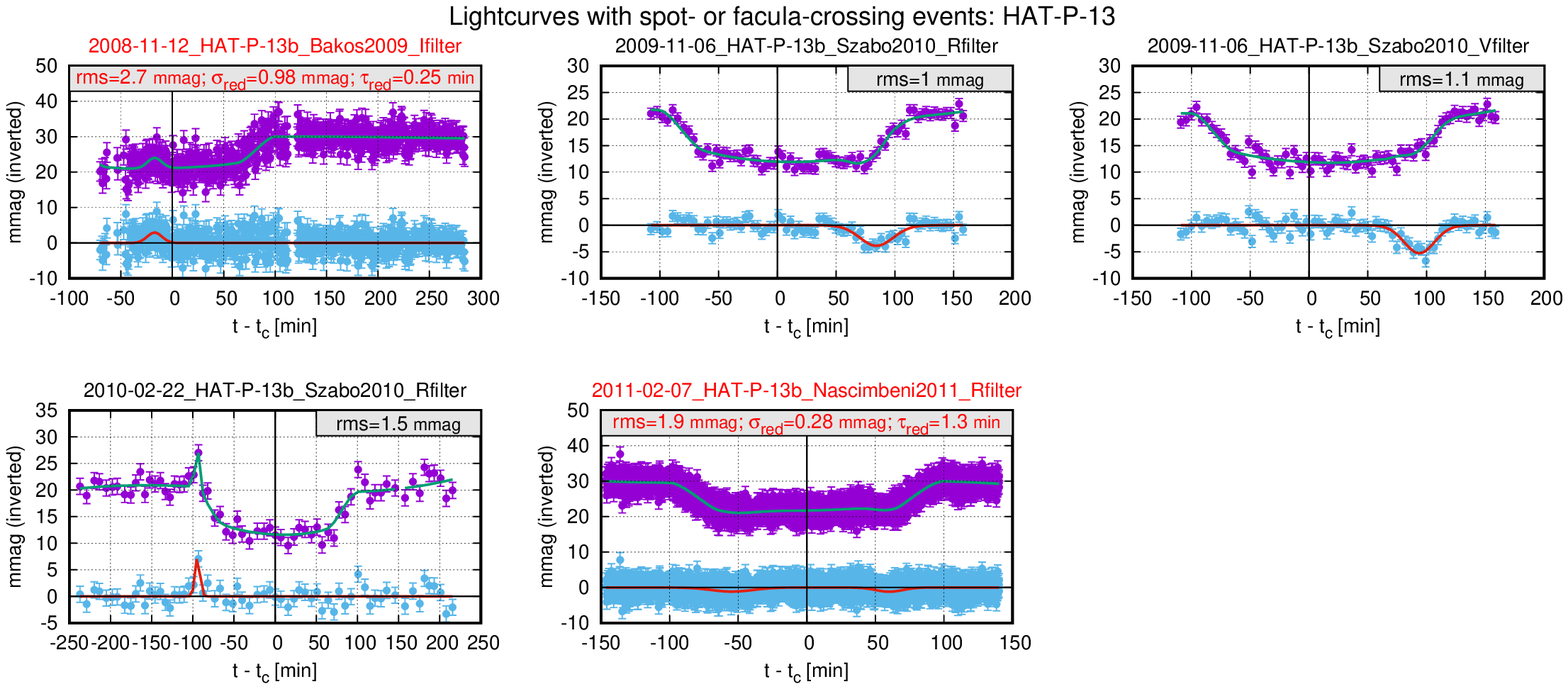}\\
\includegraphics[width=\textwidth]{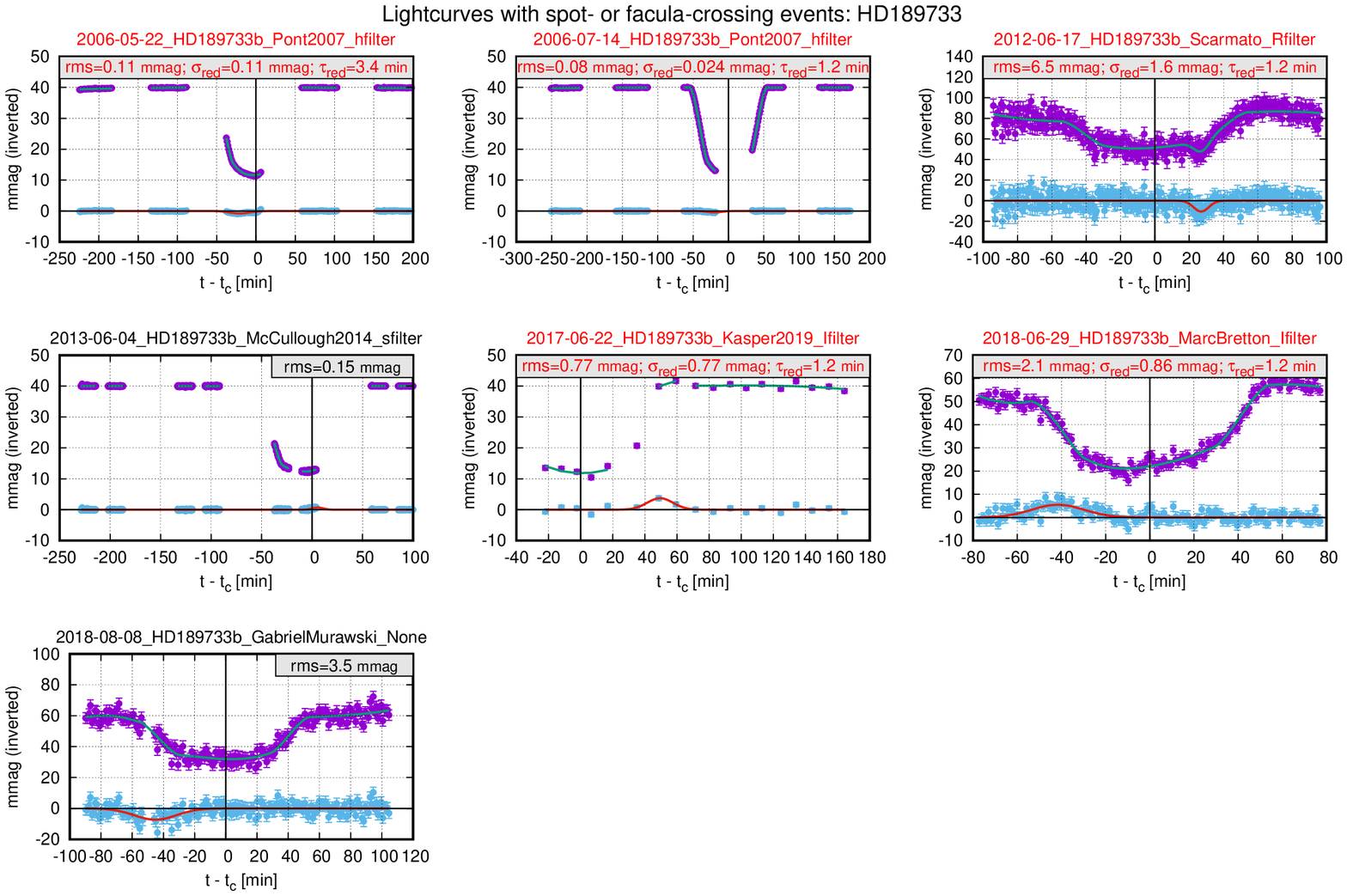}\\
\includegraphics[width=\textwidth]{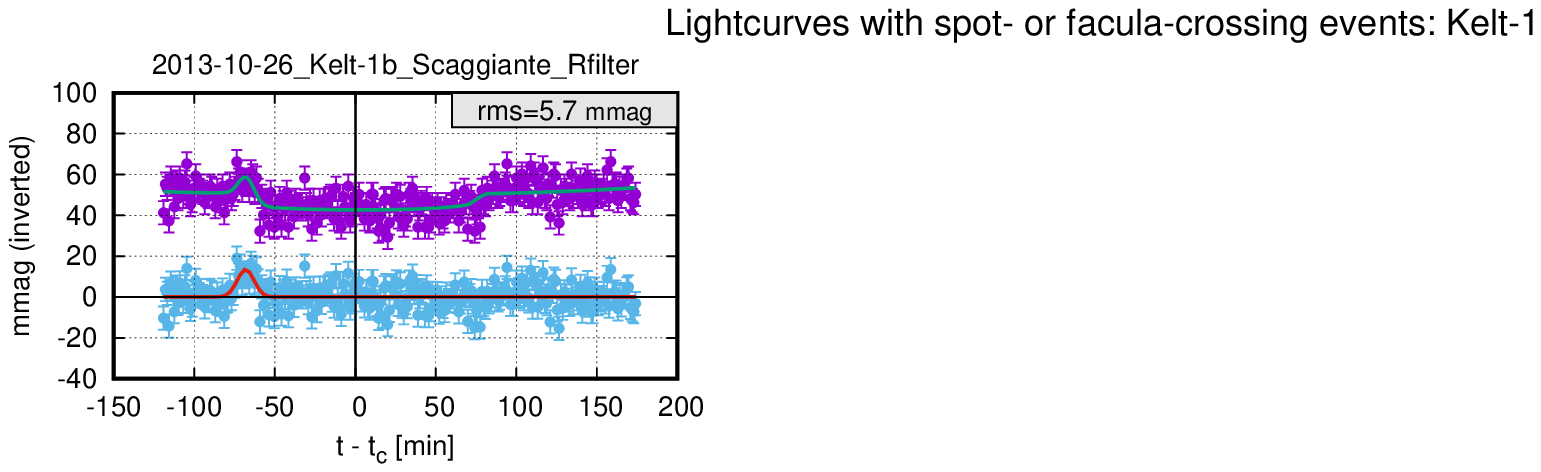}\\
\FigCap{\label{fig:spots2} GAs detected by our pipeline (Part~2: HAT-P-13, HD~189733, and Kelt-1).}
\end{figure}

\begin{figure}[t]
\includegraphics[width=\textwidth]{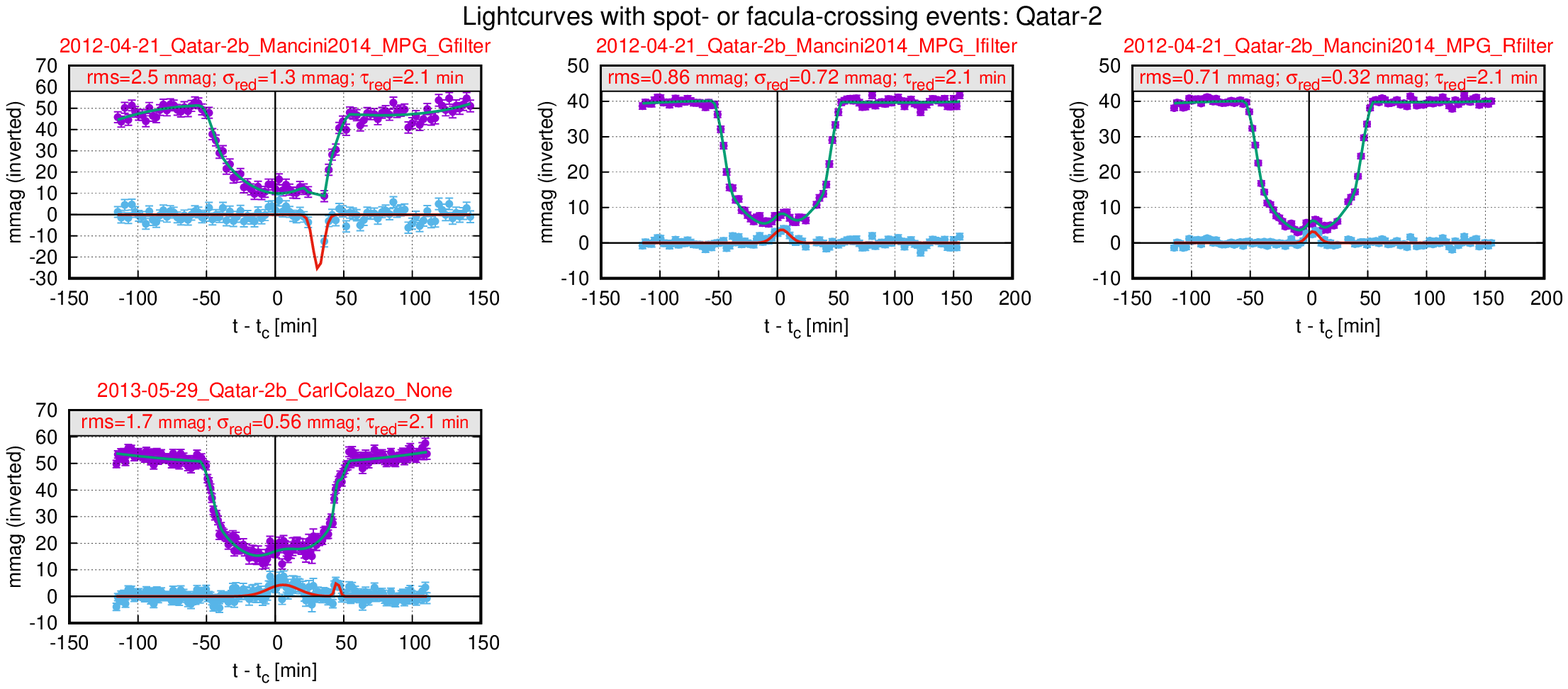}\\
\includegraphics[width=\textwidth]{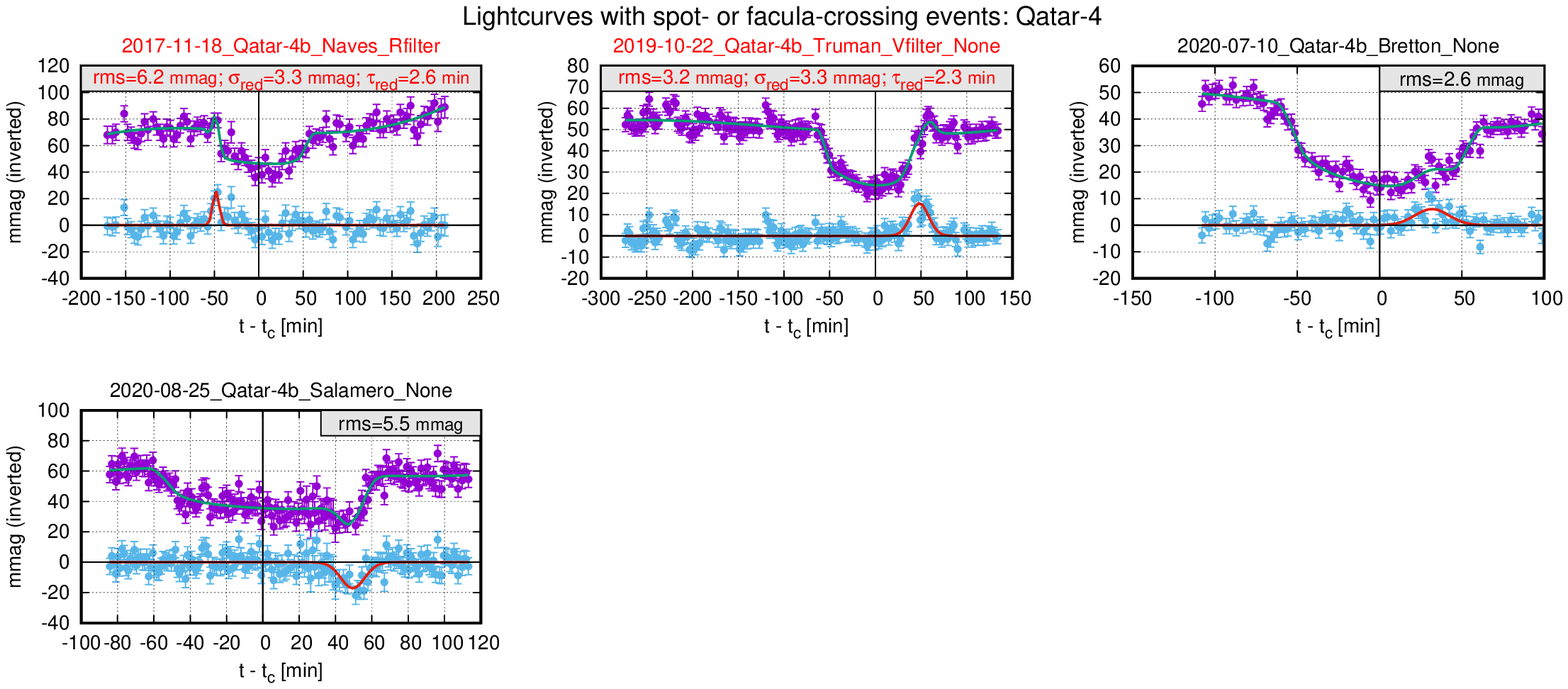}\\
\FigCap{\label{fig:spots3} GAs detected by our pipeline (Part~3: Qatar-2 and Qatar-4).}
\end{figure}

\begin{figure}[t]
\includegraphics[width=\textwidth]{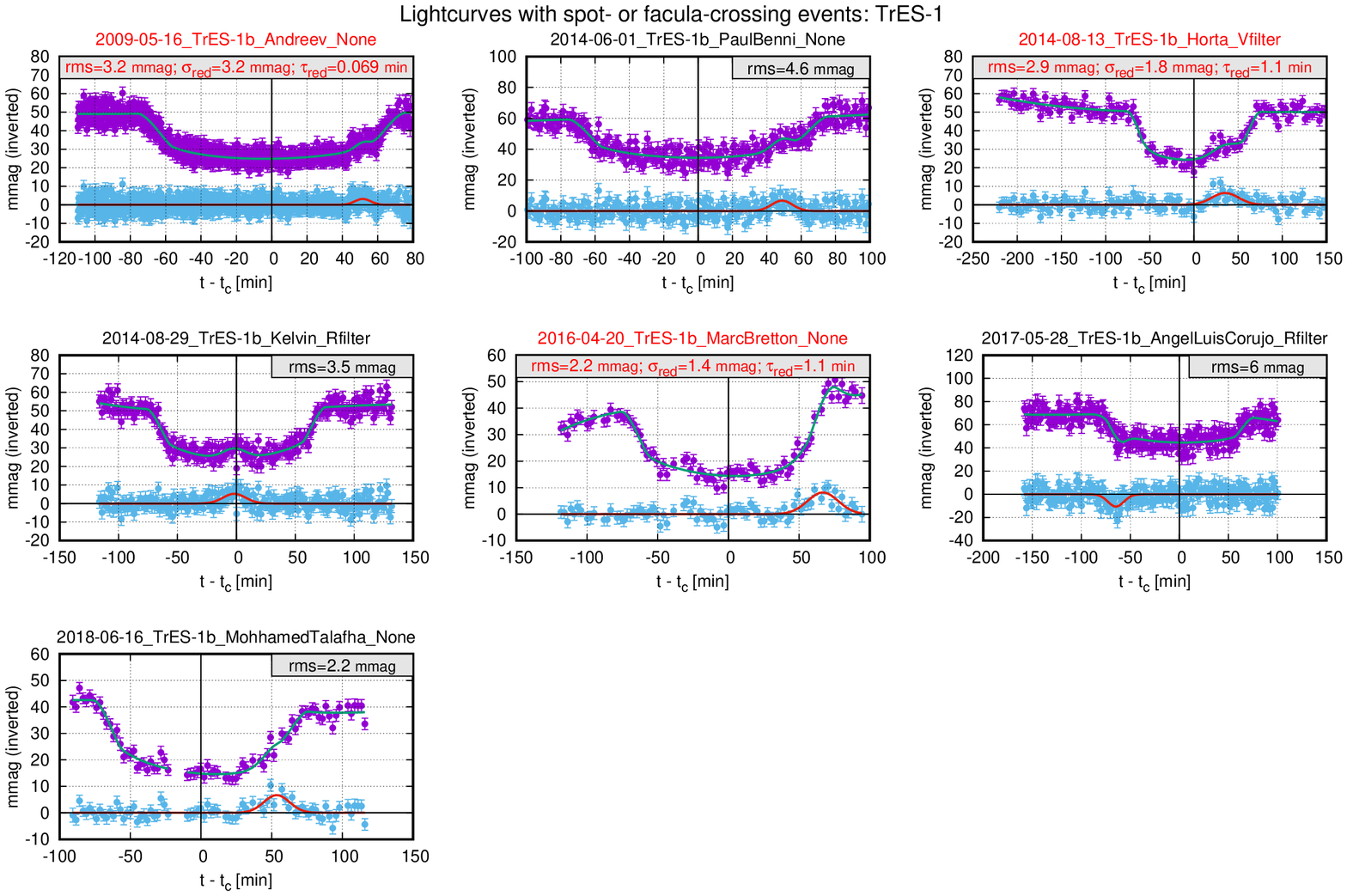}\\
\includegraphics[width=\textwidth]{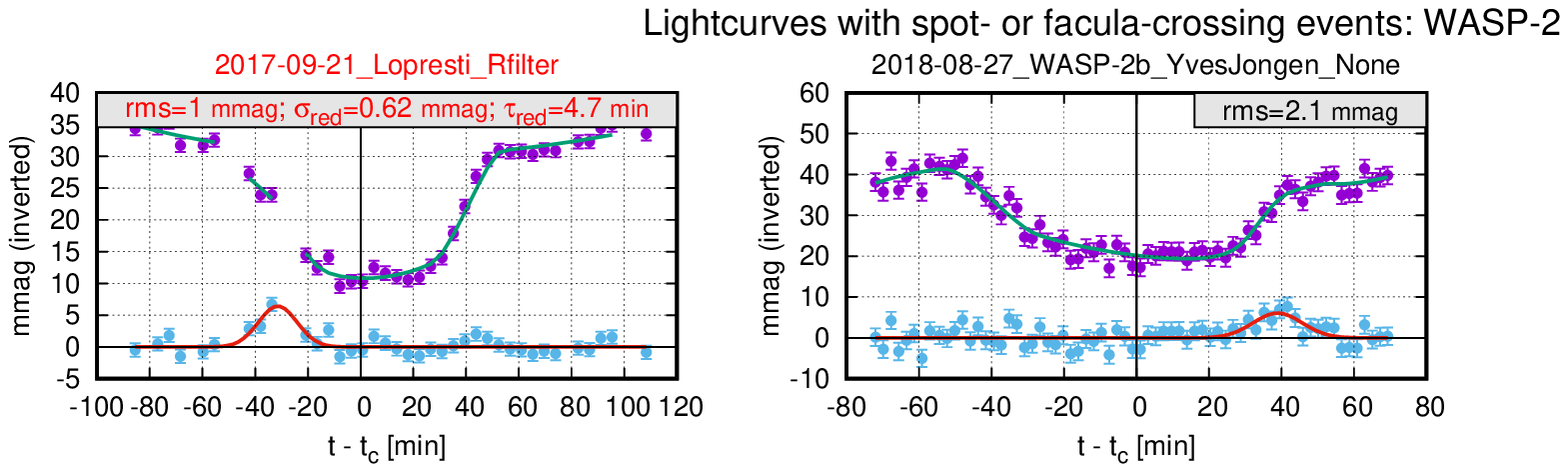}\\
\includegraphics[width=\textwidth]{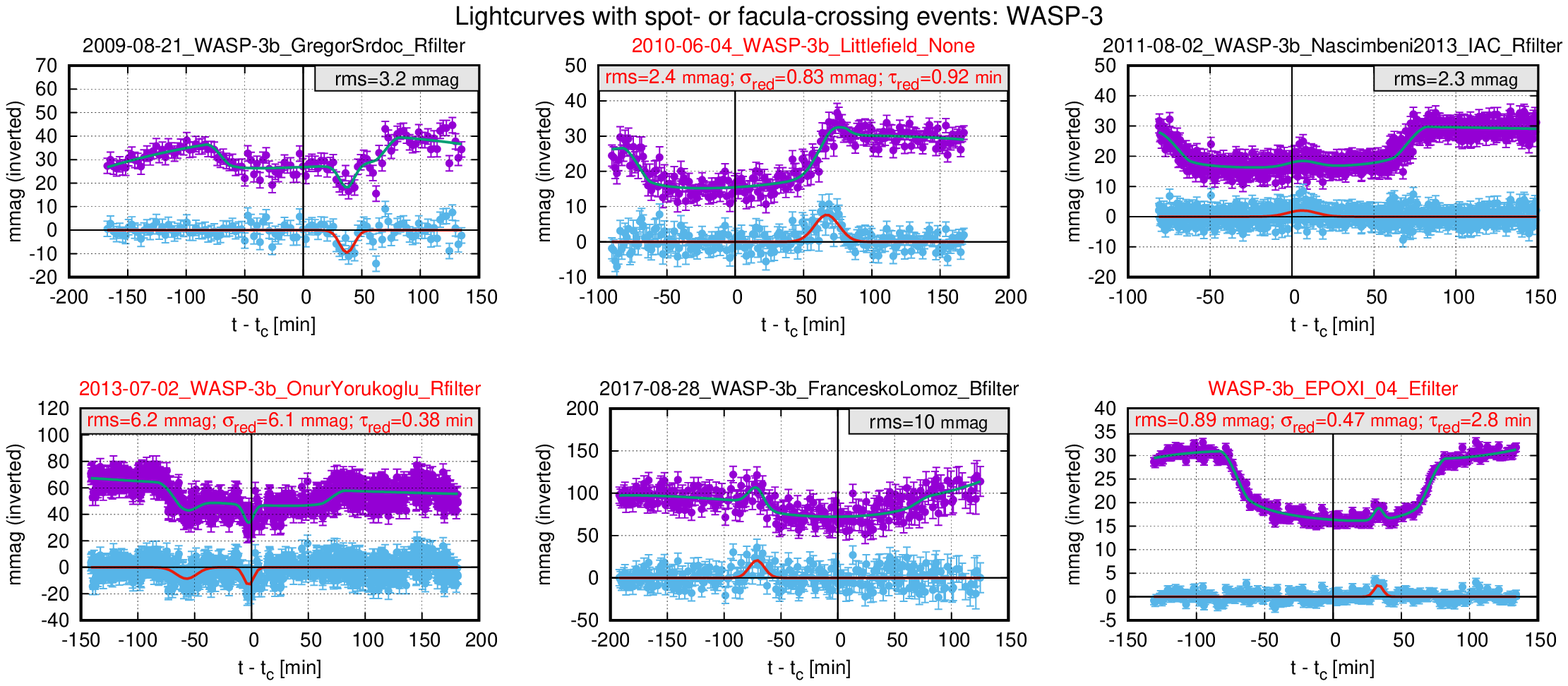}\\
\FigCap{\label{fig:spots4} GAs detected by our pipeline (Part~4: TrES-1, WASP-2, and WASP-3).}
\end{figure}

\begin{figure}[t]
\includegraphics[width=\textwidth]{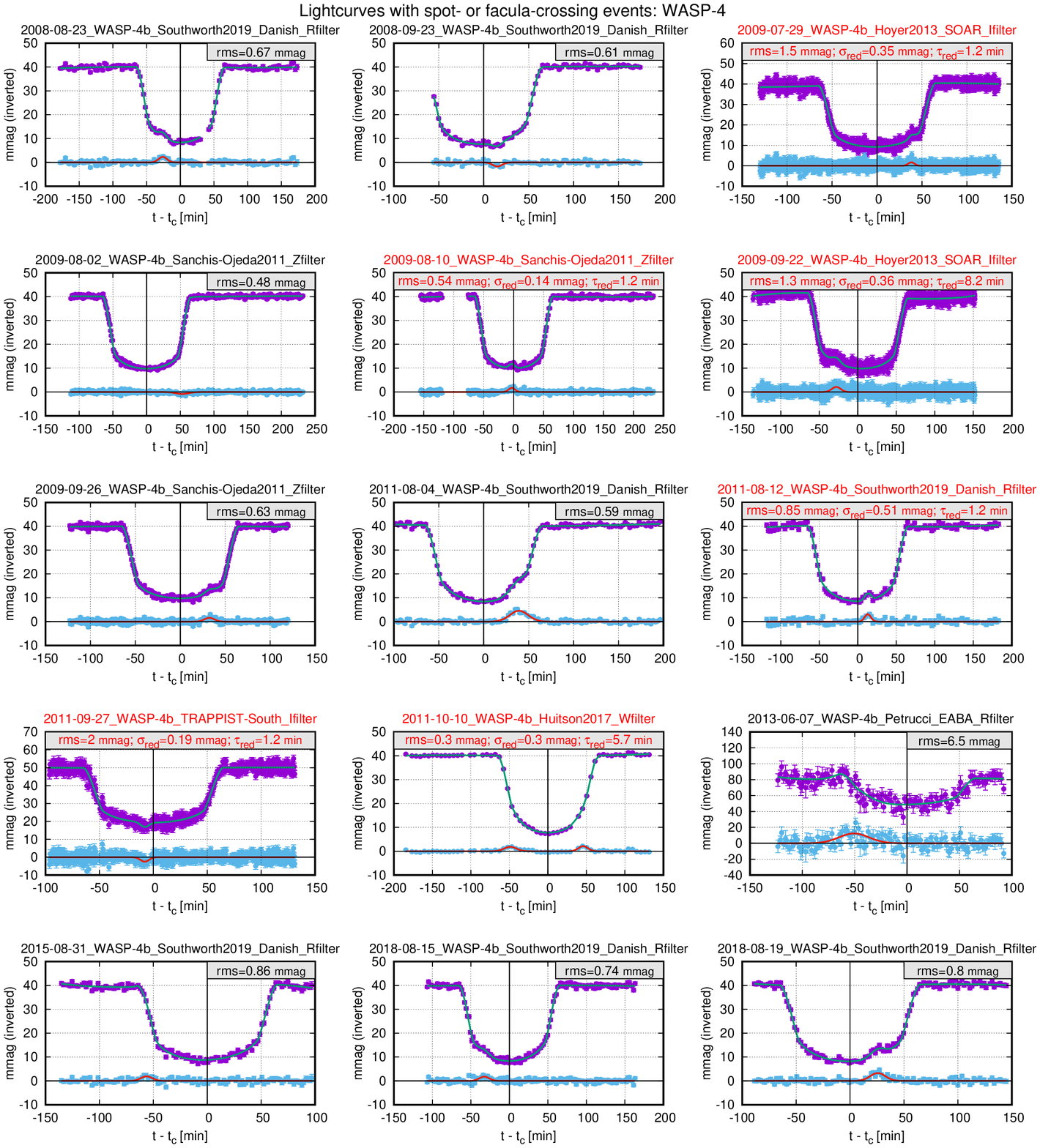}\\
\FigCap{\label{fig:spots5} GAs detected by our pipeline (Part~5: WASP-4).}
\end{figure}

\begin{figure}[t]
\includegraphics[width=\textwidth]{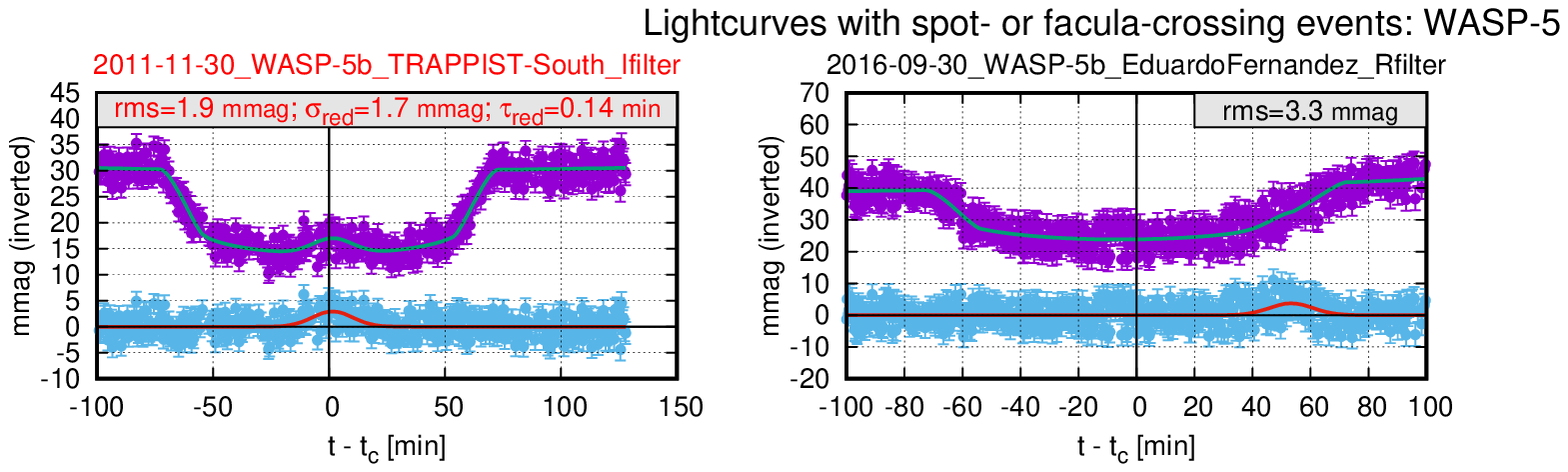}\\
\includegraphics[width=\textwidth]{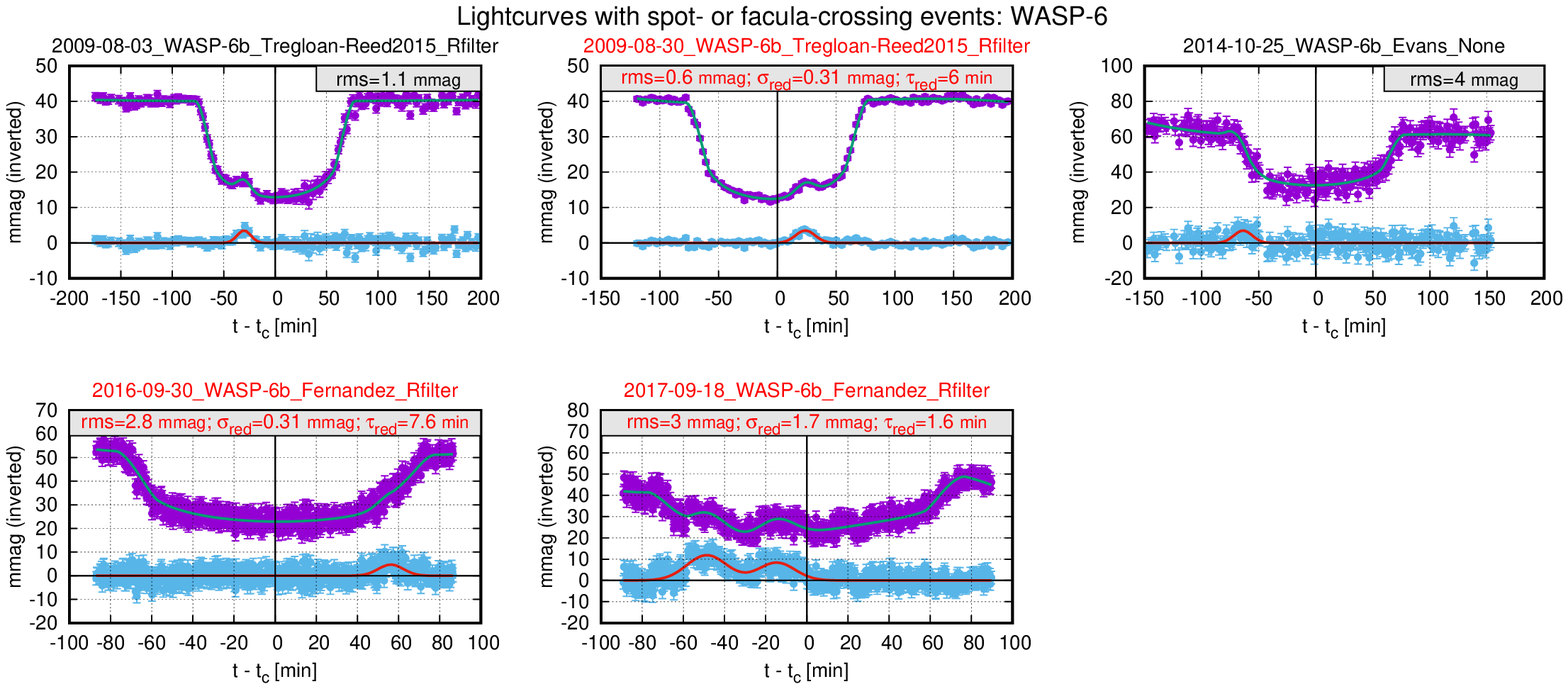}\\
\FigCap{\label{fig:spots6} GAs detected by our pipeline (Part~6: WASP-5 and WASP-6).}
\end{figure}

\begin{figure}[t]
\includegraphics[width=\textwidth]{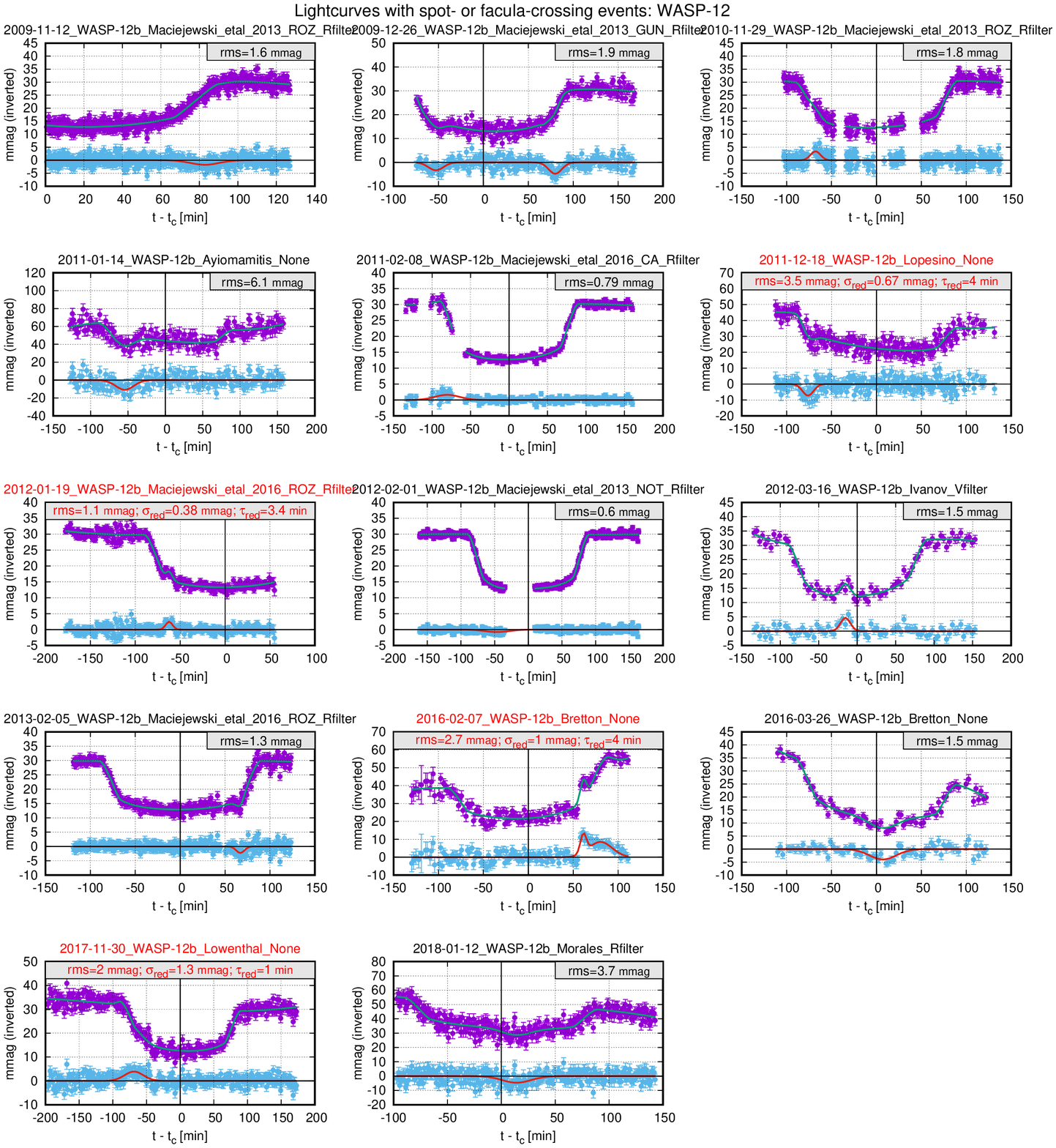}\\
\FigCap{\label{fig:spots7} GAs detected by our pipeline (Part~7: WASP-12).}
\end{figure}

\begin{figure}[t]
\includegraphics[width=\textwidth]{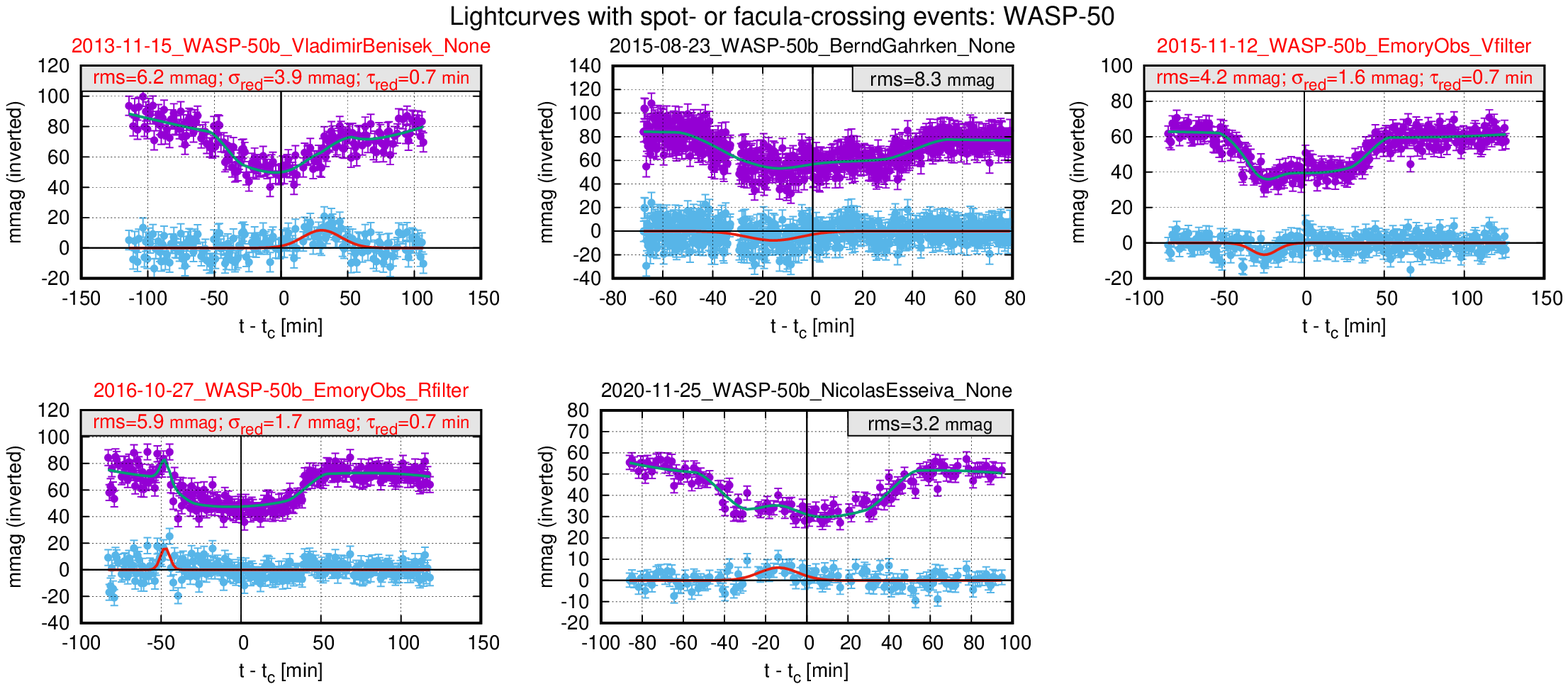}\\
\includegraphics[width=\textwidth]{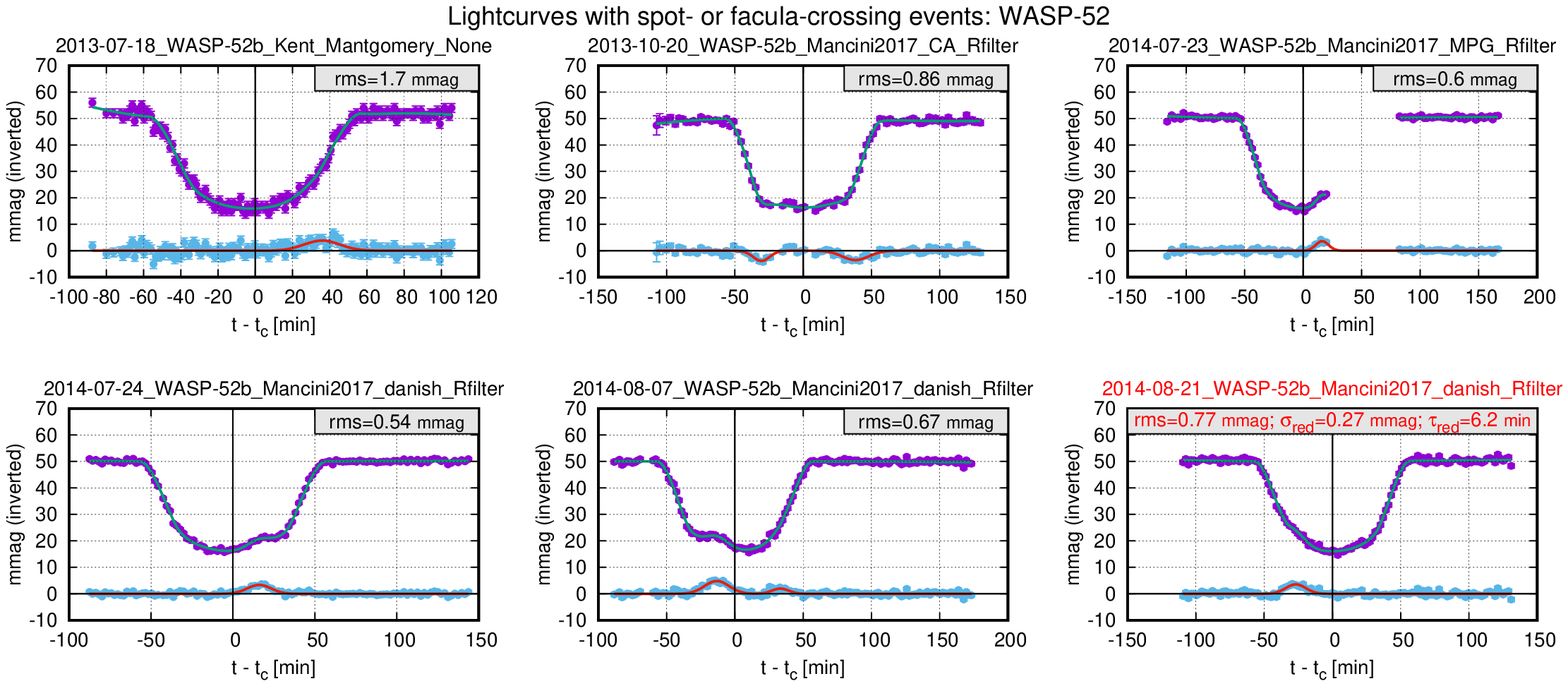}\\
\includegraphics[width=\textwidth]{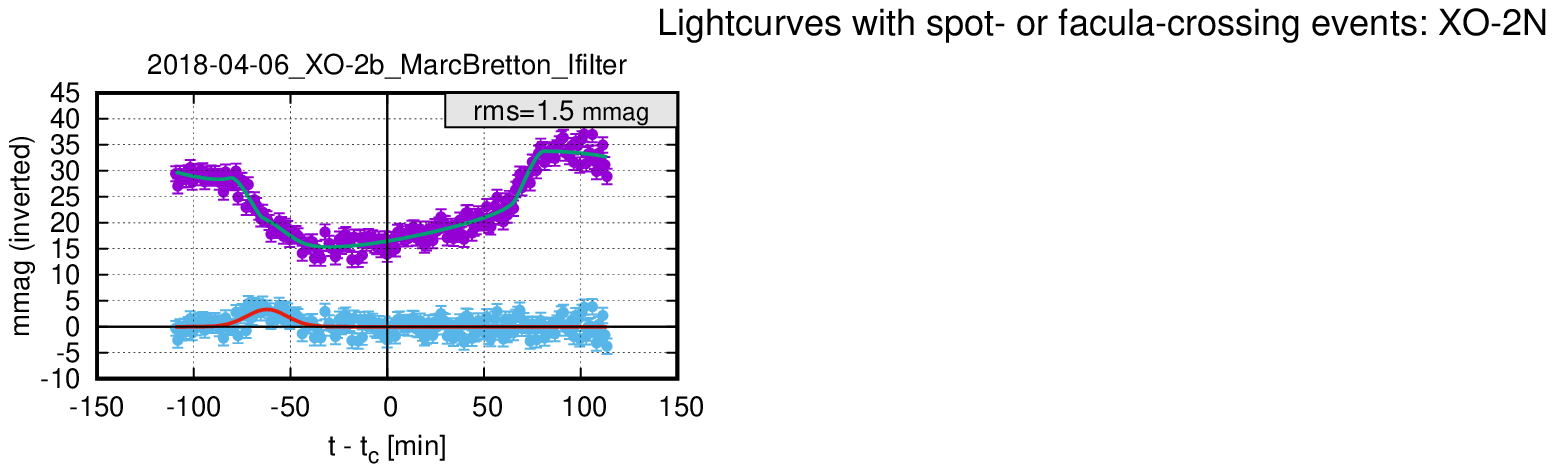}\\
\includegraphics[width=\textwidth]{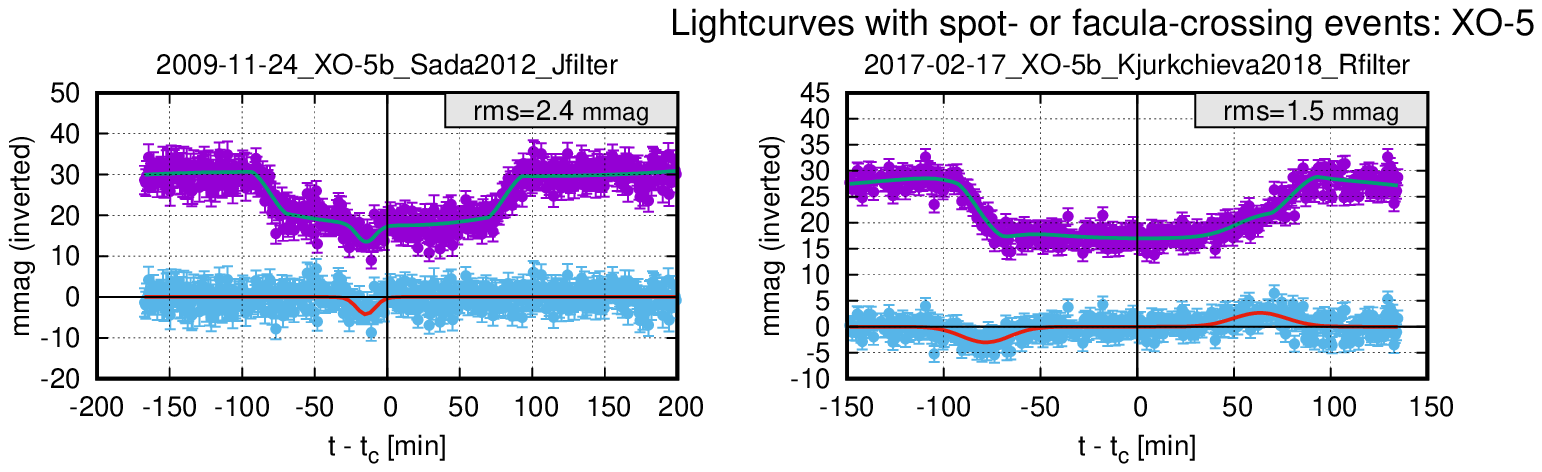}\\
\FigCap{\label{fig:spots8} GAs detected by our pipeline (Part~8: WASP-50, WASP-52, XO-2N, and XO-5).}
\end{figure}

All lightcurves with detected GAs are plotted in Figs~(\ref{fig:spots1}-\ref{fig:spots8}).
In these plots we show the original lightcurve data with their best fitting model, their
``partial'' residuals (everything subtracted except candidate GA), and model of the GA (or
multiple GAs, if present). We also print additional data in each plot, including the fit
r.m.s. and red noise parameters, if the red noise was fitted. The title is simply the file
name used in the Baluev et~al. (2019) data release. It contains the date of the
observation, target name, name(s) of the observer (or first author of a paper and a year),
and generic filter information.

The primary obvious property of our GA set is large asymmetry between positive and negative
GAs, $38$ cases against $71$. The negative ones (those that may refer to spots) are clearly
dominating. Such imbalance is difficult to explain by statistical errors, instrumental
issues, or inaccuracies of the analysis, because then the number of positive and negative
GAs should be approximately equal. Even if all $38$ positive GAs were artifacts unrelated
to stellar physics, the number of artifacts among negative GAs should be approximately the
same, so we have no less than $\sim 33$ physical spot-crossing events. Negative GAs form a
clear concentration, outlined in Fig.~\ref{fig:allspots}. Moreover, this concentration
seems slightly inclined, possibly reflecting a natural correlation with spot size (smaller
spot -- smaller GA width -- smaller GA amplitude). We fitted the logarithmic linear
regression $\log\sigma = a \log|K| + b$, restricting it to only negative GAs with $|K|$
below $\frac{1}{2}$ of transit depth (more physically reasonable cases), and we obtained
$a=0.29\pm 0.10$, a statistically remarkable value. Concerning positive GAs, we did not
detect any clear correlation and it is not obvious in Fig.~\ref{fig:allspots}.

Simultaneously, the number of non-physical GAs is likely large. Some negative GAs have
$|K|$ greater than transit depth. This is not physical, since a spot cannot have negative
brightness. If we look into particular transit curves, far not all of them reveal a
convincing GA signature. It may appear that our pipeline tried to fit some sudden noisy
spikes or instrumental events that our GP model could not predict statistically. In some
cases the residuals reveal hints of a non-stationary statistical behavior, e.g. variable
variance. Those cases are also out of our algorithm responsibility: it will approximate
such a non-stationary noise by a stationary GP model with average parameters. Some GAs
solely depend on just a single photometric observation, e.g. a Mancini et~al. (2014)
lightcurve for Qatar-2, dated by 2012-04-21 (Fig.~\ref{fig:spots3}). This point comes right
after a gap in the lightcurve, so it seems to be a deviation caused by a cloud or an
instrumental failure that was not cut away in full. As such, the GA remains unreliable even
though the suspicious observation was not classified as outlier on Stage~1. Finally, in
very accurate data like HST observations of HD~189733, it seems that we tried to fit either
inaccuracies of the quadratic limb darkening model, or an effect of imperfect detrending,
through GAs.

Many such cases have to remain inconclusive, because their classification cannot be
performed based on just a single photometric curve. Although several transits in our
database were observed from independent sites, this usually did not appear helpful enough,
because the photometric accuracy varies for different sites.\footnote{For example, the
WASP-4 transit lightcurve on 23.08.2008 by (Southworth et~al., 2019) reveals a
spot-crossing event. This transit was simultaneously observed by Hoyer et~al. (2013) from
two telescopes, but those two lightcurves appeared too noisy for a robust verification.
Nearly the same story is about 23.09.2008 facula-crossing event detected in (Southworth
et~al., 2019) data. See Fig.~\ref{fig:spots5}.} To clearly ensure that a particular GA is a
physical crossing event rather than noise artifact it is necessary to have a complex
same-high-quality multi-site and multi-channel (including e.g. spectral) observations of a
single transit. This would be too expensive programme perhaps, but in view of our results
it is most important to seek such comprehensive characterization for
\emph{positive} GAs that indicate faculae-crossings.

\section{Discussion}
We developed an algorithmic pipeline that allows to perform massive detection of GAs that
possibly refer to spot- and facula-crossing events in transit lightcurves. Although the
algorithm was based on a statistically rigorous mathematics, it relies on particular
models: Gaussian model of the anomaly, quadratic limb-darkening law, exponential
correlation function of the noise, and a stationary GP model. Any of these models may turn
inadequate for a particular lightcurve, or it may involve statistically unpredictable
instrumental or weather events. Because of all these factors, our algorithm collects weird
systematic patterns together with physical spot- and facula-crossing. Therefore, it is
necessary to develop additional post-filtering criteria rejecting unreliable GAs (e.g.
those that depend on just a single measurement, or which involve sudden lightcurve jumps,
or the noise demonstrates non-stationary behavior).

However, we can already conclude that our pipeline is pretty efficient in what concerns the
automated detection of spot anomalies in good data. For example, for WASP-4 we detected $4$
of $6$ spot-transit events found by Southworth et~al. (2019). One their proposed
spot-crossing did not pass our reliability test (it appeared too wide) and the other one
appeared statistically insignificant, but in exchange we detected two other spot-crossings
and one facula-crossing in their data. We believe this agreement is good enough, and
simultaneously our method has more solid math grounds in comparison with visual detection
approach by Southworth et~al. (2019).

One of our underlying goals was to improve transit timing accuracy by modeling spot
anomalies. As noticed in (Baluev et~al., 2019), there is typically an excess jitter in
measured transit times, and this effect clearly depends on the star. For example, HD~189733
revealed a significant TTV jitter of $\sim 1.5$~min beyond uncertainties, likely indicating
its larger activity in comparison with other targets. Our hypothesis was that by performing
massive spot detection and modeling, this TTV jitter can be removed. However, for HD~189733
we detected approximately an average relative number of GAs, and many of them did not look
convincingly robust after all. The resulting effect on TTV variance seems rather
negligible, because only a few per cent of lightcurves revealed GAs. Simultaneously, we
detected many apparently reliable spots for targets like WASP-4, WASP-52, and WASP-12,
which all have paradoxically small TTV jitter. It seems that the activity-related TTV
jitter comes from another physical phenomena, not directly related to detectable spots.

Yet another issue comes from ambiguous interpretation of spot anomalies. Too often the
models ``transit+spot'' and ``shifted transit'' appear statistically indistinguishable. In
fact, most of the moderate shifts of transit mid-times can be equally explained by spot
anomalies in the ingress and/or egress phases. An example is the single WASP-4 Huitson
et~al. (2017) lightcurve where we detected two ``side'' spots
(Fig.~\ref{fig:spots5}).\footnote{Such paired GAs may also be caused by an inaccuracy of
the limb-darkening model, however it seems unlikely in this particular case, because the
limb darkening was determined based on all four Huitson et~al. (2017) lightcurves. So the
issue is that one of these lightcurves, shown in the plot, has a signficantly different
shape than three others in average.} If we did not fix its mid-time at a quadratic
ephemeris, we would likely select a more simple lightcurve model with only one of these GAs
with roughly doubled amplitude. This would result in a shifted timing, either positive or
negative, depending on which GA we discard. Such cases trigger ambiguous interpretation of
the data and bi-modal timing estimates. Time series of this type, with bi-modal
measurements, are quite unusual and their analysis needs a better understanding.

\Acknow{Organization of the EXPANSION project (ENS, IAS, GGV), development of data
processing algorithms (RVB, VShSh) and statistical analysis (RVB) were supported by the
Russian Science Foundation, project~19-72-10023. Collecting the database of third-party
exoplanetary data (RVB) and part of EXPANSION observations (VNA, GShM, AFV, DRG and GGV)
were supported by the Ministry of Science and Higher Education of Russian Federation,
project 075-15-2020-780 (manuscript Section~\ref{sec_pipe}) AGG and GMB acknowledge
``Fundamental Research'' government contract of the Special Astrophysical Observatory of
the Russian Academy of Sciences and the Federal program ``Kazan Federal University
Competitive Growth Program''. EP acknowledges the Europlanet 2024 RI project funded by the
European Union's Horizon 2020 Research and Innovation Programme (Grant agreement No.
871149). TCH acknowledges financial support from the National Research Foundation (NRF; No.
2019R1I1A1A01059609). TRAPPIST is funded by the Belgian Fund for Scientific Research (Fond
National de la Recherche Scientifique, FNRS) under the grant FRFC~2.5.594.09.F, with the
participation of the Swiss National Science Fundation (SNF). MG and EJ are F.R.S.-FNRS
Senior Research Associates. TRAPPIST data sets analyzed in this study are available at the
ESO Science Archive Facility at~\url{http://archive.eso.org/}. We express gratitude to the
anonymous reviewer of the manuscript for their useful comments.}

\end{document}